\documentclass[useAMS,usegraphicx,usenatbib]{mn2e}

\title[The Nearly Universal Merger Rate]{The Nearly Universal Merger Rate
   of Dark Matter Haloes in $\Lambda$CDM Cosmology}

\author[O Fakhouri and C-P Ma]{Onsi Fakhouri$^{1}$\thanks{E-mail:
    onsi@berkeley.edu, cpma@berkeley.edu} and  Chung-Pei Ma$^{1}$\\
  $^{1}$Department of Astronomy, 601 Campbell Hall, University of
  California, Berkeley, CA 94720}

\pagerange{\pageref{firstpage}--\pageref{lastpage}}

\usepackage{amsmath}
\usepackage{multirow}
\newcommand{\mfof}{M_{FOF}}
\newcommand{\mvir}{M_{200}}
\newcommand{\mt}{z_{P}\!\!:\!\!z_{D}}
\newcommand{\dz}{\Delta z}
\newcommand{\Reqn}{\frac{d\bar{N}_{\rm merge}}{dt}}
\newcommand{\Rzeqn}{\frac{d\bar{N}_{\rm merge}}{dz}}
\newcommand{\R}{d\bar{N}_{\rm merge}/dt}
\newcommand{\Rz}{d\bar{N}_{\rm merge}/dz}

\newcommand{\mmin}{2\times10^{12}M_\odot}

\begin{document}

\label{firstpage}

\maketitle

\begin{abstract}
  We construct merger trees from the largest database of dark matter haloes
  to date provided by the Millennium simulation to quantify the merger
  rates of haloes over a broad range of descendant halo mass ($10^{12} \la
  M_0 \la 10^{15} M_\odot$), progenitor mass ratio ($10^{-3} \la \xi \le
  1$), and redshift ($0 \le z \la 6$). We find the mean merger rate {\it
    per halo}, $B/n$, to have very simple dependence on $M_0$, $\xi$, and
  $z$, and propose a universal fitting form for $B/n$ that is accurate to
  10-20\%. Overall, $B/n$ depends very weakly on the halo mass ($\propto
  M_0^{0.08}$) and scales as a power law in the progenitor mass ratio
  ($\propto \xi^{-2}$) for minor mergers ($\xi \la 0.1$) with a mild upturn
  for major mergers. As a function of time, we find the merger rate per Gyr
  to evolve roughly as $(1+z)^{n_m}$ with $n_m=2-2.3$, while the rate per unit
  redshift is nearly independent of $z$. Several tests are performed to
  assess how our merger rates are affected by, e.g. the time interval
  between Millennium outputs, binary vs multiple progenitor mergers, and
  mass conservation and diffuse accretion during mergers. In particular, we
  find halo fragmentations to be a general issue in merger tree
  construction from $N$-body simulations and compare two methods for
  handling these events.  We compare our results with predictions of two
  analytical models for halo mergers based on the Extended Press-Schechter
  (EPS) model and the coagulation theory.  We find that the EPS model
  overpredicts the major merger rates and underpredicts the minor merger
  rates by up to a factor of a few.
\end{abstract}

\section{Introduction}

In hierarchical cosmological models such as $\Lambda$CDM, galaxies'
host dark matter haloes grow in mass and size primarily through mergers
with other haloes.  As the haloes merge, their more centrally concentrated
baryonic components sink through dynamical friction and merge subsequently.
The growth of stellar masses depends on both the amount of mass brought in
by mergers and the star formation rates.  Having an accurate description of
the mergers of dark matter haloes is therefore a key first step in
quantifying the mergers of galaxies and in understanding galaxy formation
and growth.

Earlier theoretical studies of galaxy formation typically relied on merger
trees generated from Monte Carlo realisations of the merger rates given by
the analytical extended Press-Schechter (EPS; \citealt{1993MNRAS.262..627L,
  1991ApJ...379..440B}) model (e.g. \citealt{1993MNRAS.264..201K,
  1999MNRAS.310.1087S, 2000MNRAS.319..168C}).  Some recent studies have
chosen to bypass the uncertainties and inconsistencies in the EPS model by
using halo merger trees from $N$-body simulations directly
(\citealt{1999MNRAS.303..188K, 2000MNRAS.311..793B, 2003MNRAS.338..903H,
  2005ApJ...631...21K, 2005Natur.435..629S}).  As we find in this paper,
obtaining robust halo merger rates and merger trees requires rich halo
statistics from very large cosmological simulations as well as careful
treatments of systematic effects due to different algorithms used for,
e.g., assigning halo masses, constructing merger trees, removing halo
fragmentation events, and choosing time spacings between simulation
outputs.

The aim of this paper is to determine the merger rates of dark matter
haloes as a function of halo mass, merger mass ratio (i.e. minor vs major),
and redshift, using numerical simulations of the $\Lambda$CDM cosmology.
This basic quantity has not been thoroughly investigated until now mainly
because large catalogues of haloes from finely spaced simulation outputs
are required to provide sufficient merger event statistics for a reliable
construction of merger trees over a wide dynamic range in time and mass.
We achieve this goal by using the public database of the Millennium
simulation \citep{2005Natur.435..629S}, which follows the evolution of
roughly $2\times10^{7}$ dark matter haloes from redshift $z=127$ to $z=0$.
This dataset allows us to determine the merger rates of dark matter haloes
ranging from galaxy-mass scales of $\sim 10^{12} M_\odot$ over redshifts
$z=0$ to $\sim 6$, to cluster-mass scales up to $\sim 10^{15} M_\odot$ for
$z=0$ to a few.  We are also able to quantify the merger rates as a
function of the progenitor mass ratio $\xi$, from major mergers ($\xi \ga
0.1$) down to minor mergers of $\xi \sim 0.03$ for galaxy haloes and down
to $\xi \sim 3\times 10^{-4}$ for cluster haloes.

The inputs needed for measuring merger rates in simulations include a
catalogue of dark matter haloes and their masses at each redshift, and
detailed information about their ancestry across redshifts, that is, the
merger tree.  Unfortunately there is not a unique way to identify haloes,
assign halo masses, and construct merger trees.  In this paper we primarily
consider a halo mass definition based on the standard friends-of-friends
(FOF) algorithm and briefly compare it with an alternative mass definition
based on spherical overdensity.

For the merger trees, we investigate two possible algorithms for treating
events in which the particles in a given progenitor halo end up in more
than one descendant halo ('fragmentations').  We find that these events are
common enough that a careful treatment is needed.  In the conventional
algorithm used in the literature, the progenitor halo is linked one-to-one
to the descendant halo that has inherited the largest number of the
progenitor's particles.  The ancestry links to the other descendant haloes
are severed (for this reason we call this scheme 'snipping').  We consider
an alternative algorithm ('stitching') in this paper, in which fragmentations
are assumed to be artefacts of the FOF halo identification scheme.  We
therefore choose to recombine the halo fragments and stitch them back into
the original FOF halo.

Earlier theoretical papers on merger rates either relied on a small sample
of main haloes to estimate the overall redshift evolution over a limited
range of halo masses, or were primarily concerned with the mergers of {\it
  galaxies} or {\it subhaloes}.  For halo mergers, for example,
\cite{1999AJ....117.1651G} studied $z < 1$ major mergers of
galaxy-sized haloes in an open CDM and a tilted $\Omega_m=1$ CDM model using
$N$-body simulations in a 100 Mpc box and $144^3$ particles. 
\cite{2001ApJ...546..223G} used a sample of $\sim 4000$ haloes to study the
environmental dependence of the redshift evolution of the major merger rate
at $z < 2$ in $\Lambda$CDM.  \cite{2006ApJ...652...56B} studied major
mergers of subhaloes in $N$-body simulations in a 171 Mpc box with $512^3$
particles and the connection to the observed close pair counts of galaxies.
For galaxy merger rates, \cite{2002ApJ...571....1M} and
\cite{2006ApJ...647..763M} are based on up to $\sim 500$ galaxies formed
in SPH simulations in $\sim 50$ Mpc boxes with up to $144^3$ gas particles, while
\cite{2007arXiv0708.1814G} used the semi-analytical galaxy catalogue of
\cite{2006MNRAS.366..499D} based on the Millennium simulation.

This paper is organised as follows. Section~\ref{MillenniumSection}
describes the dark matter haloes in the Millennium simulation
(\S\ref{DarkMatterHaloes}) and how we construct the merger trees
(\S\ref{mergertrees}) .  We then discuss the issue of halo fragmentation
and the two methods ('snipping' and 'stitching') used to treat these
events in \S\ref{FOFLims}.  The notation used in this paper is summarised
in \S\ref{Notation}.

Section~\ref{StatsSection} describes how mergers are counted
(\S\ref{MultiCount}) and presents four (related) statistical measures of
the merger rate (\S\ref{stats}). The relation between these merger rate
statistics and the analytical merger rate based on the Extended
Press-Schechter (EPS) model is derived in Section~\ref{EPSConnection}.

Our main results on the merger rates computed from the Millennium
simulation are presented in Section~\ref{Results}.  We first discuss the
$z\approx 0$ results and quantify the merger rates as a function of the
descendant halo mass and the progenitor mass ratios using merger trees
constructed from the stitching method (\S\ref{MRz0}).  The evolution of
the merger rates with redshifts up to $z\sim 6$ is discussed in
Section~\ref{redshiftdependence}.  We find a simple universal form for the
merger rates and present an analytic fitting form that provides a good
approximation (at the 10-20\% level) over a wide range of parameters
(\S\ref{FitSection}).

Section~\ref{Tests} compares the stitching and snipping merger rates
(\S\ref{SnipvsStitch}) and presents the key results from a number of tests
that we have carried out to assess the robustness of our results.  Among
the tests are: time convergence and the dependence of the merger rates on
the redshift spacing $\dz$ between the Millennium outputs used to construct
the merger tree (\S\ref{TimeResolutionSection}); how the counting of binary
vs multiple progenitor mergers affects the merger rates
(\S\ref{MultiCountValid}); mass non-conservation arising from 'diffuse'
accretion in the form of unresolved haloes during mergers
(\S\ref{MassCons}); and how the definition of halo masses and the treatment
of fragmentation events affect the resulting halo mass function
(\S\ref{MassFunction}).

In Section~\ref{DiscussionSection}, we discuss two theoretical frameworks
that can be used to model halo mergers: EPS and coagulation.  A direct
comparison of our merger rates and the EPS predictions for the Millennium
$\Lambda$CDM model shows significant differences over a large range of
parameter space (\S\ref{eps}).  Section~\ref{Coagulation} discusses
Smoluchowski's coagulation equation and the connection between our merger
rates and the coagulation merger kernel.

The appendix compares a third merger tree (besides snipping and stitching)
constructed from the Millennium catalogue by the Durham group
\citep{2006MNRAS.370..645B, 2006MNRAS.367.1039H,2003MNRAS.338..903H}.  Two
additional criteria are imposed on the subhaloes in this algorithm to
reduce spurious linkings of FOF haloes.  We find these criteria to result
in reductions in both the major merger rates and the halo mass function.

The cosmology used throughout this paper is identical to that used in the
Millennium simulation: a $\Lambda\textrm{CDM}$ model with $\Omega_m=0.25$,
$\Omega_b=0.045$, $\Omega_\Lambda=0.75$, $h=0.73$, an initial power-law
index $n=1$, and $\sigma_8=0.9$ \citep{2005Natur.435..629S}.  Masses and
lengths are quoted in units of $M_\odot$ and Mpc without the Hubble
parameter $h$.

\section{Haloes and Merger Trees in the Millennium Simulation}
\label{MillenniumSection}

\subsection{Dark Matter Haloes}\label{DarkMatterHaloes} 

The Millennium simulation provides the largest database to date for
studying the merger histories of dark matter haloes in the
$\Lambda\mbox{CDM}$ cosmology.  The simulation uses $2160^{3}$ particles
with a particle mass of $1.2\times10^{9}M_{\odot}$ in a $685$ Mpc box and
traces the evolution of roughly $2\times10^{7}$ dark matter haloes from
redshift $z=127$ to $z=0$ \citep{2005Natur.435..629S}.

The haloes in the simulation are identified by grouping the simulation
particles using the standard friends-of-friends algorithm (FOF:
\citealt{1985ApJ...292..371D}) with a linking length of $b=0.2$.
Each FOF halo (henceforth referred to as \emph{FOF} or \emph{halo}) is then
broken into constituent subhaloes by the SUBFIND algorithm, which
identifies dark matter substructure as locally overdense regions within
each FOF and removes any remaining gravitationally unbound particles
\citep{2001MNRAS.328..726S}.  The result is a list of disjoint subhaloes
typically dominated by one large background host subhalo and a number of
smaller satellite subhaloes.

Each subhalo in the catalogue is assigned a mass given by the number of
particles bound to the subhalo; only subhaloes with more than 20 simulation
particles are included in the database.  Each FOF halo is then given two
definitions of mass: $\mfof$, which counts the number of particles
associated with the FOF group, and $\mvir$, which assumes the halo is
spherical and computes the virial mass within the radius at which the
average interior density of the halo is 200 times the mean density of the
universe.  $\mfof$ includes background particles that are unbound by the
SUBFIND algorithm so it is generally larger than the sum of the subhalo
masses.  In this paper we mainly use $\mfof$ as it is found to be the more
robust mass definition in our merger study.  We discuss $\mvir$ and a
number of mass conservation issues in Section~\ref{MassCons}.

\begin{figure}
\centering
\includegraphics{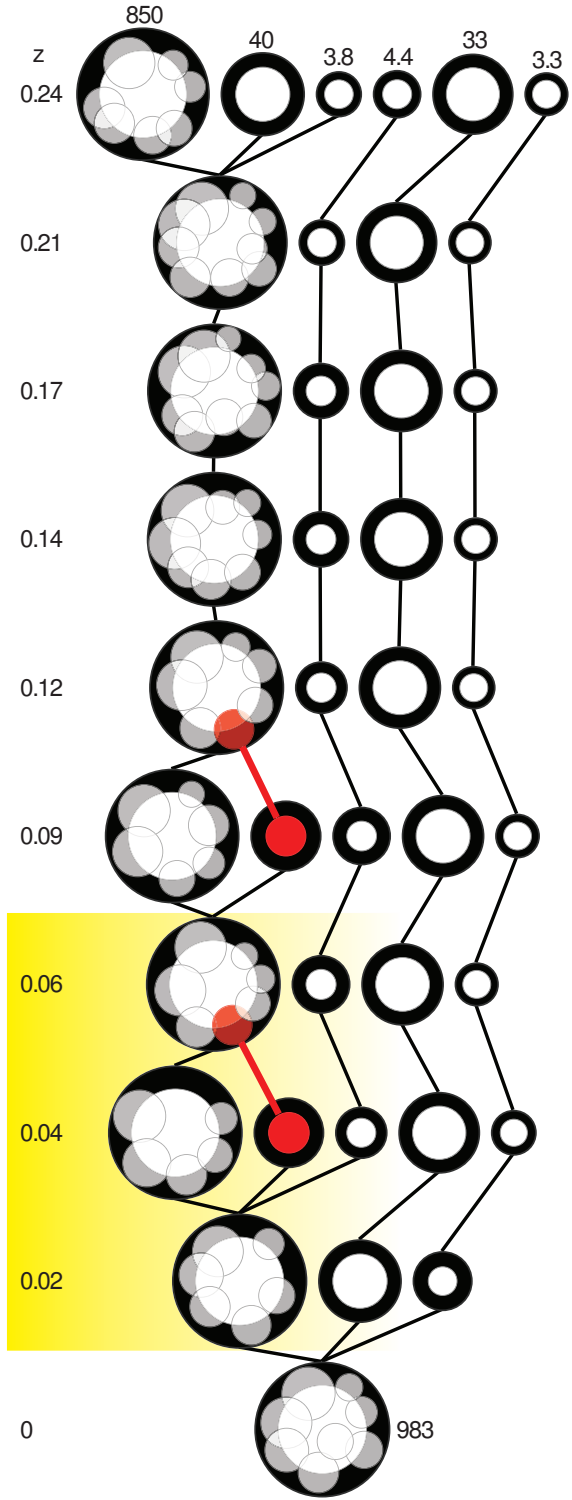}
\caption{Example of a typical FOF merger tree extracted from the Millennium
  database. Black circles denote FOF haloes; white circles within black
  circles denote subhaloes.  The radius of each circle is proportional to
  the log of the mass of the object; the black circles are further scaled
  up by a factor of 1.5 for clarity.  (The locations of the white circles
  within their parent FOF haloes are drawn randomly.)  Red circles denote
  fragmenting subhaloes. The highlighted (yellow) fragmentation event is
  studied in Fig. \ref{fig:fragmentation}.  The numbers above the haloes at
  $z=0.24$ and to the right of the final descendant FOF at $z=0$ correspond
  to the FOF masses (in units of $10^{10} M_\odot$).  }
\label{fig:mergertree}
\end{figure}

\begin{figure}
\centering
\includegraphics{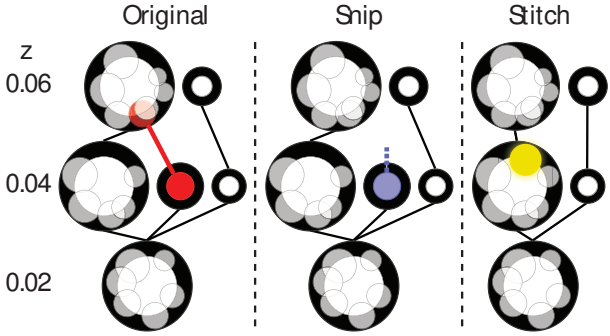}
\caption{Left: A closeup of the highlighted (yellow region) fragmentation
  event in Fig.~\ref{fig:mergertree}.  The middle and right panels
  illustrate how the snipping and stitching methods handle fragmentation in
  order to assign a unique descendant halo.  The blue circle (centre panel)
  shows the snipped orphan subhalo, and the yellow circle (right panel)
  shows how that subhalo is stitched.  The black, white, and red circles are
  the same as in
  Fig.~\ref{fig:mergertree}.  }
\label{fig:fragmentation}
\end{figure}

\subsection{Merger Tree Construction}
\label{mergertrees}

Merger trees of dark matter haloes in the Millennium database are
constructed by connecting \emph{subhaloes} (\emph{not the FOF haloes})
across 64 snapshot outputs: a subhalo at a given output is taken to be the
\emph{descendant}\footnote{It is common practice in the literature to call
  the descendant halo the \emph{parent} halo even though the parent is
  formed later and, hence, is younger than the progenitor.  We avoid this
  confusing notation throughout.} of a \emph{progenitor} subhalo at a prior
output (i.e. higher redshift) if it contains the largest number of bound
particles in the progenitor subhalo.  This procedure results in a merger
tree in which each progenitor subhalo has a single descendant subhalo, even
though in general, the particles in the progenitor do not necessarily all
end up in the same descendant subhalo.

It is worth noting that merger trees in $N$-body simulations are typically
constructed based on the FOF haloes and not on the subhaloes.  The standard
way of assigning the progenitor and descendant FOF haloes in those studies,
however, is the same as the procedure applied to the {\it subhaloes} in
Millennium discussed above; that is, the descendant halo is the halo that
inherits the most number of bound particles of the progenitor.  As will be
elaborated on below, we call this the 'snipping' method.

The focus of this paper is on the merger history of the \emph{FOF} haloes
rather than the subhaloes, so we must process the subhalo merger tree
available from the public database to construct a consistent merger tree
for the FOF haloes.  We consider an FOF halo A to be a descendant of an
earlier FOF halo B if B contains a subhalo whose descendant subhalo is in
A.  Progenitor FOF haloes are said to have merged when all their descendant
subhaloes are identified with one descendant FOF.  We illustrate this
process in Fig.~\ref{fig:mergertree} with an actual merger tree taken from
the Millennium database.  The upper left corner, for example, shows three
FOF haloes at $z=0.24$ with masses $8.5\times 10^{12}$, $4\times 10^{11}$,
and $3.8\times 10^{10} M_\odot$ merging into a single FOF halo at the next
Millennium output ($z=0.21$).  The largest FOF halo $z=0.24$ has 7
subhaloes (white circles) in addition to the host (sub)halo, while each of
the two smaller FOF haloes has only one host (sub)halo.  For clarity, the
ancestral links between subhaloes are suppressed in
Fig.~\ref{fig:mergertree}.

\begin{figure}
\centering
\includegraphics{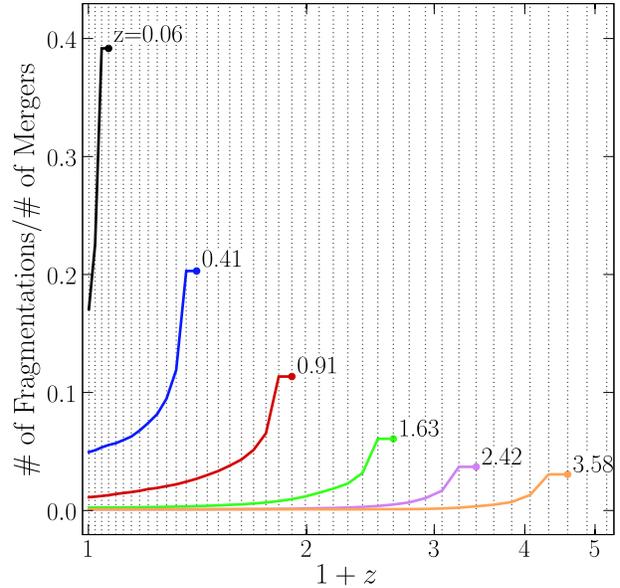}
\caption{Distribution of the ratio of fragmentation to merger events as a
  function of redshift. The dotted vertical lines correspond to the
  redshifts of the Millennium outputs. We choose 6 redshifts (labelled) for
  illustrative purposes and plot the ratio of the number of fragmentations
  to the number of mergers (filled circles) at each redshift.  A mass ratio
  cutoff is applied: both the fragments and mergers must have mass ratios
  exceeding 10\%.  The line emanating from each circle then traces the
  evolution of the number of fragmentation events (the number of mergers
  being held fixed), which drops as subhalo fragments remerge with their
  original FOF halo.  We note that about half of the subhalo fragments
  remerge within 2-3 simulation outputs.  Finally, the six filled circles
  decrease with increasing redshift, reaching $\sim 40$\% at $z=0$ but
  dropping to $\sim 5$\% at high $z$ -- this is primarily due to the
  increasing $\dz$ between Millennium outputs.}
\label{fig:fragdist}
\end{figure}

\subsection{Halo Fragmentation} \label{FOFLims}

Even though each subhalo in the Millennium tree, by construction, is
identified with a single descendant subhalo (see last subsection), the
resulting FOF tree can have fragmentation events in which an FOF halo is
split into two (or more) descendant FOF haloes.  The red circles in
Fig.~\ref{fig:mergertree} at $z=(0.12:0.09)$ and (0.06:0.04) illustrate two
such events: the subhaloes of the progenitor FOF halo end up in different
descendant FOF haloes.  It is important to emphasise that this
fragmentation issue is not unique to the use of subhaloes in the Millennium
simulation, but rather occurs in general in any merger tree construction
where groups of particles at two different redshifts must be connected.
This is because particles in a progenitor halo rarely end up in exactly one
descendant halo; a decision must therefore be made to select a unique
descendant.  There is not a unique way to do this, and we explore below two
methods that we name \emph{snipping} and \emph{stitching} to handle these
fragmentation events.

Fig.~\ref{fig:fragmentation} illustrates these two methods for the
fragmentation event shown in the highlighted (yellow) region of
Fig.~\ref{fig:mergertree}.  The snipping method is commonly used in the
literature (e.g. \citealt{2002MNRAS.329...61S}), presumably for its
simplicity.  Fragmentation events are removed by 'snipping' the link
between the smaller descendant halo and its progenitor FOF halo, as shown
in the middle panel of Fig.~\ref{fig:fragmentation}.  The fragmenting
progenitor FOF halo then has only one descendant FOF halo.  We note that
this method can result in a number of progenitor-less \emph{orphan} FOF
haloes (e.g., the blue subhalo in Fig.~\ref{fig:fragmentation}).

In this paper we investigate an alternative method that we name
'stitching.'  This method is motivated by our observation that about half
of the fragmented haloes in the Millennium simulation remerge within the
following 2-3 outputs (see below).  The two fragmentation events in
Fig.~\ref{fig:mergertree} both belong to this category: the fragmented
haloes at $z=0.09$ and 0.04 (red circles) are seen to have remerged by the
following output time ($z=0.06$ and 0.02).  This behaviour is not too
surprising because merging haloes oscillate in and out of their respective
virial radii before dynamical friction brings them into virial equilibrium
(typically on timescales of a few Gyrs; see, e.g.,
\citealt{2008MNRAS.383...93B}).  During this merging phase, the FOF halo
finder can repeatedly disassociate and associate the progenitor haloes,
leading to spurious fragmentation and remerger events and inflating the
merger rate.  This behaviour needs to be taken into account before a robust
merger rate can be obtained.

We therefore do not count remerging fragments as merger events in the
``stitching'' method.  Specifically, we group the fragmented haloes into
two categories: those that remerge within 3 outputs after fragmentation
occurs, and those that do not.  The fragmented haloes that remerge are
stitched into a single FOF descendant (e.g. the yellow subhalo in the right
panel of Fig.~\ref{fig:fragmentation}); those that do not remerge are
snipped and become orphan haloes. Often the fragment subhaloes have become
members of a new FOF group that is otherwise unrelated to the original FOF.
In such instances they are removed from that group and stitched into the
main FOF descendant\footnote{There is, however, one exceptional case: if a
  subhalo fragment becomes the largest subhalo of an FOF, all subhaloes in
  that FOF are stitched into the fragment's original FOF.}.  A further test
of the dependence of our results on the choice of 3 outputs is described in
Section~\ref{SnipvsStitch}.

As can be seen in Fig.~\ref{fig:fragmentation}, the snipping method will
yield a higher merger rate than stitching due to the remerger events.  We
quantify the relative importance of these events in
Fig.~\ref{fig:fragdist}, where the ratio of fragmentation events to merger
events is seen to peak at 40\% for major fragmentation events (defined to
be fragmentations where the fragment subhalo carries 10\% or more of the
halo mass) at low-$z$ and falls off at high $z$ where $\dz$ is large.  For
the fragmentation events occurring at a given redshift $z_f$ in
Fig.~\ref{fig:fragdist} (filled circles), the drop of each curve with
decreasing $z$ tracks how many of them have remerged by that redshift.  As
noted above, we find that about half of the fragmented haloes remerge
within 2-3 outputs (corresponding to a fixed $\dz/(1+z)$ as the outputs are
log-spaced).  Given a fragmentation-to-merger ratio of 40\%, and a remerger
rate of 50\%, the remerging fragments can impact the merger rate
measurements inflating them at the $\sim$20\% level.

Moreover, we find that this effect is more severe for fragmentations where
the mass of the fragment is small relative to the mass of the original
parent halo (we call these minor fragmentations).  If we consider
fragmentations in which the subhalo fragments carry between 1\% and 10\% of
the original FOF mass, the fragmentation-to-merger ratio at $z=0$ ($z=1.6$)
jumps to 57\% (13.3\%) vs 39\% (6\%) for major fragmentations.  For very
minor fragmentations (subhalo fragments that carry less than 1\% of the
total mass) the fragmentation-to-merger ratios are 85\% and 28\% at $z=0$
and $z=1.6$ respectively.  Thus we anticipate that fragmentation events
will more severely pollute the minor-merger regime of the merger rate
statistics.

\subsection{Notation}
\label{Notation}

We apply both the stitching and snipping methods and produce FOF merger
trees from the 46 Millennium outputs that span $z=0$ and $z=6.2$.  From
these trees we connect different outputs and generate a catalogue of
descendant FOF haloes at the low-z ($z_D$) output and their associated
progenitor FOF haloes at the high-z ($z_P$) output.  We refer to this as
the $\mt$ catalogue and produce a number of catalogues for a variety of
output spacings.  The redshift spacing is denoted by $\dz=z_P-z_D$.  The
Millennium outputs are logarithmically distributed, providing fine $\dz$
down to 0.02 (corresponding to $\sim$260 Myrs) near $z=0$ and larger $\dz$
at high redshifts, e.g., $\dz \approx 0.1$ at $z\approx 1$ and $\approx
0.5$ at $z\approx 6$.  Specifically, the lowest 10 redshift outputs are at
0.0, 0.02, 0.04, 0.06, 0.09, 0.12, 0.14, 0.17, 0.21, and 0.24.

\begin{table*}
\centering
\begin{tabular}{l|ccc|ccc|ccc}
	\multirow{2}{*}{$\mt$} & \multicolumn{3}{c}{Galaxy-Scale} & \multicolumn{3}{c}{Group-Scale} & \multicolumn{3}{c}{Cluster-Scale} \\
	&$N_p=1$&$N_p=2$&$N_p>2$&$N_p=1$&$N_p=2$&$N_p>2$&$N_p=1$&$N_p=2$&$N_p>2$ \\
	\hline
	0.06:0  & 188,400 & 65,711 & 27,939  & 1,063 & 2,418 & 13,256  & 3 & 25 & 5,356\\
	0.56:0.51  & 189,351 & 61,718 & 22,031  & 1,212 & 2,468 & 9,374  & 6 & 18 & 3,014 \\
	1.08:0.99  & 145,779 & 68,467 & 35,426  & 325 & 878 & 7,630  & 0 & 2 & 1,308 \\
	2.07:1.91  & 76,298 & 52,525 & 39,097  & 31 & 77 & 2,225  & 0 & 0 & 129 \\
	3.06:2.83  & 30,641 & 26,675 & 25,072  & 0 & 4 & 343  & 0 & 0 & 4 \\

\end{tabular}
\caption{
  The number of merger events in the Millennium simulation
  that we use to determine the merger rates.
  Merger trees at five representative redshifts are shown: $z\approx 0$, 0.5, 1, 2, and 3.
  At each $z$, we list the number of FOF haloes that have a single
  progenitor halo ($N_p=1$, i.e., no mergers), two progenitors ($N_p=2$, i.e.
  binary mergers), and multiple progenitors ($N_p>2$),
  for three separate descendant mass bins:
  $2\times10^{12}\leq M_0 < 3\times10^{13} M_\odot$ (galaxy), $3\times10^{13}\leq M_0
  < 10^{14} M_\odot$ (group), and $M_0 \geq 10^{14}M_\odot$ (cluster).
  Only progenitor haloes with mass $> 4.8\times 10^{10} M_\odot$ (40 simulation particles) are counted.
}
\label{table:NumProgs}
\end{table*}

For a given FOF descendant halo in a $\mt$ catalogue, we use $M_{0}$ to
denote its $\mfof$ mass, $N_p$ to denote the number of progenitor haloes,
and $M_{i}$ with $i\in (1,2,\dots,N_{p})$ to denote the rank-ordered
$\mfof$ mass of the progenitors, i.e. $M_1\geq M_2\geq \dots M_{N_p}$.  We
impose a minimum mass cutoff of $M_{0}\geq\mmin$ on the descendant FOF halo
and a cutoff of $M_i\geq 4.8\times10^{10}M_\odot$ on the progenitors, which
corresponds to 40 particles and is twice the minimum halo mass in the
Millennium database.

For certain results reported below, we make use of three large mass bins:
$2\times10^{12}\leq M_0 < 3\times10^{13} M_\odot$, $3\times10^{13}\leq M_0
< 10^{14} M_\odot$, and $10^{14}M_\odot\leq M_0$, referred to as the
\emph{galaxy-scale}, \emph{group-scale}, and \emph{cluster-scale} bins,
respectively.

%%%%%%%%%%%%%%%%%%%%%%%%%%%%%%%%%%%%%%%
\begin{figure}
\centering
\includegraphics{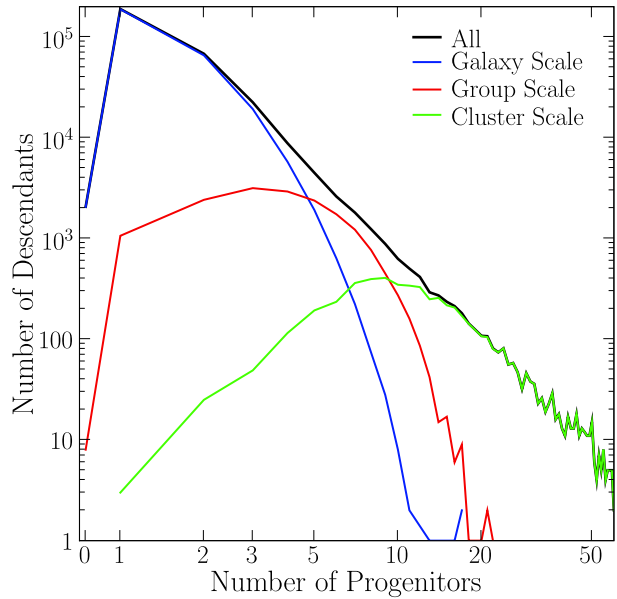}
\caption{Distribution of the number of progenitors, $N_p$, for the
  $z$ = 0.06:0 merger tree. There are $\sim300,000$ descendant FOF haloes at
  redshift 0 (black) with $M_0\geq\mmin$.  Of these $\sim280,000$ have
  $2\times10^{12}\leq M_0 < 3\times10^{13} M_\odot$ (galaxy-scale; dark
  blue), $\sim16,000$ have $3\times10^{13}\leq M_0 < 10^{14} M_\odot$
  (group scale; red), and $\sim 5,400$ have $M_0 \geq 10^{14}M_\odot$
  (cluster-scale; green).}
\label{fig:NumProgs}
\end{figure}
%%%%%%%%%%%%%%%%%%%%%%%%%%%%%%%%%%%%%%%

\section{Merger Statistics and Connection to EPS} 
\label{StatsSection}

\subsection{Counting Many-to-One Mergers} \label{MultiCount}

Despite the fine time spacing between Millennium's outputs, a
non-negligible number of the descendant FOF haloes have more than two
progenitors listed in the merger tree (i.e. $N_p>2$).
For completeness, we list in Table~\ref{table:NumProgs} the actual number
of merger events in the Millennium simulation available to us after we
construct the FOF merger trees.  Statistics at five representative
redshifts are shown: $z\approx 0$, 0.5, 1, 2, and 3.  At each $z$, we list
separately the number of FOF haloes that have $N_p=1, 2$ and $>2$
progenitor haloes, for three separate descendant mass bins.  As expected
for hierarchical cosmological models, the halo numbers drop with increasing
$z$ and increasing $M_0$.

Fig.~\ref{fig:NumProgs} shows the distribution of the number of
progenitors, $f(N_p)$, for the $z$ = 0.06:0 merger tree for the same three
mass bins.  Only the stitching method is shown; the snipping method has a
similar distribution.  We find that $(62, 22, 16)$\% of the haloes have
$N_p=(1,2,>2)$ identifiable progenitors at $z=0.06$; more than half of the
FOF haloes at $z=0$ therefore have only one progenitor at $z=0.06$ and did
not experience a merger during this redshift interval.  When separated into
different descendant mass bins, the peak of $f(N_p)$ moves to higher $N_p$
for more massive haloes.  For a fixed $(z_P,z_D)$, clusters therefore tend
to have more progenitors, and unlike galaxy-mass haloes, very few of the
cluster haloes are single-progenitor events (i.e. $N_p=1$)

For completeness, we include {\em all} the progenitors (above our minimum
mass cutoff of 40 particles) in our merger rate statistics.  Since we have
no information about the order in which the multiple progenitors merge with
one another, we assume that each progenitor halo $M_i$ with $i\geq 2$
merges with $M_1$, the most massive progenitor, at some stage between the
two outputs.  Thus a descendant halo with $N_p$ progenitors is assumed to
be the result of a sequence of $(N_p-1)$ binary merger events, where each
merger event is assigned a mass ratio
\begin{equation}
     \xi \equiv \frac{M_i}{M_1} \,, \quad i=2,...,N_p
\end{equation}
which by construction satisfies $\xi\leq1$.  This assumption ignores the
possibility that two smaller progenitor FOF haloes merge together before
merging with the most massive progenitor.  Section~\ref{MultiCountValid}
describes how we have tested the validity of this assumption and found
negligible effects as long as a sufficiently small $\dz$ is used.

\subsection{Definitions of Merger Rates} \label{stats}

In this subsection we define four related quantities that will be used to
measure the merger rates of dark matter haloes.  Merger rates can be
measured in either per Gyr or per unit redshift; the two sets of quantities
are related by a factor of $dt/dz$.  We will present most of our results in
units of per redshift since, as we will show below, the merger rates have a
particularly simple form in those units.

As a starting point, we consider the symmetric merger rate
\begin{equation}
  B_{MM'}(M, M', \mt) dM dM' \,,
\label{B1}
\end{equation}
which measures the mean merger rate (i.e. the number of mergers per unit
redshift) per unit volume between progenitor FOF haloes in the mass range
($M$,$M+dM$) and ($M'$, $M'+dM'$).  We compute this quantity using merger
trees constructed between the progenitor output redshift $z_P$ and the
descendant output redshift $z_D$.  Note that $B_{MM'}(M,M')$ has units of
$\left[\textrm{number of
    mergers}\times\,(\dz)^{-1}\,\textrm{Mpc}^{-3}\,M_\odot^{-2}\right]$ and
generally depends on both $z_P$ and $z_D$.

Instead of the individual progenitor masses $M$ and $M'$, it is often
useful to express merger rates as a function of the {\em descendant} FOF
mass and the mass ratio of the progenitors.  We do this by transforming
$B_{MM'}(M,M')dM dM'$ to
\begin{equation}
	B(M_0,\xi,\mt) dM_0 d\xi \,,
\label{B2}
\end{equation}
which measures the mean merger rate (per volume) for descendant FOF haloes
in the mass range ($M_0$, $M_0+dM_0$) at redshift $z_{D}$ that have
progenitor FOF haloes at $z_P$ with mass ratio in the range of ($\xi$,
$\xi+d\xi$), where $\xi=M_i/M_1, i\geq2$ as discussed in
Section~\ref{MultiCount}.  The quantity $B(M_0,\xi)$ therefore has units of
$\left[\textrm{number of
    mergers}\times\,\dz^{-1}\,\textrm{Mpc}^{-3}\,M_\odot^{-1}\,d\xi^{-1}\right]$.
In the mass-conserving binary limit of $M_0=M+M'$ and $\xi=M'/M$ (where
$M'<M$), $B_{MM'}$ and $B$ in equations~(\ref{B1}) and (\ref{B2}) are
related by a simple transformation.  In practice, the relation between the
two quantities is complicated by multiple mergers and imperfect merger mass
conservation.

Since the halo abundance in a $\Lambda$CDM universe decreases with
increasing halo mass, many more haloes contribute to the merger rates in
equations~(\ref{B1}) and (\ref{B2}) in the lower mass bins of $M$, $M'$, or
$M_0$.  It is useful to normalise out this effect and calculate the mean
merger rates {\it per halo}.  To do this, we divide out the number density
of the descendant FOF haloes from the merger rate $B$ and define:
\begin{equation}
	\frac{B}{n} \equiv \frac{B(M_0,\xi,\mt)}{n(M_0,z_D)} \,,
\label{B3}
\end{equation}
which measures the mean number of mergers \emph{per halo} per unit redshift
for a descendant halo of mass $M_0$ with progenitor mass ratio $\xi$; the
units are $\left[\textrm{number of mergers}/\textrm{number of
    descendants}\times\,(\dz)^{-1}\,(d\xi)^{-1}\right]$, which is
dimensionless.  The mass function $n(M_0,z)dM_0$ gives the number density
of the descendant FOF haloes with mass in the range of $(M_0,M_0+dM_0)$.

The differential merger rates defined above can be integrated over $\xi$
and $M_0$ to give the mean merger rate over a certain range of merger mass ratios
for haloes in a given mass range.  Explicitly, the mean rate of mergers for
descendant haloes in mass range $M_0 \in [m, M]$ with progenitor mass
ratios in the range $\xi \in (x,X)$,
\begin{equation}
	\Rzeqn\left([m,M],[x,X],\mt\right)  \,,
\end{equation}
is simply an integral over $B(M_0,\xi,\mt)$:
\begin{equation}
\Rzeqn\equiv\frac{1}{N}\int_{m}^{M}\!\!\!\!\int_{x}^{X}\!\!B(M_0,\xi,\mt)\,d\xi\,dM_0\,, 
\label{eqn:mergerrate}
\end{equation}
where
\begin{equation}
	N\equiv\int_{m}^{M}\!\!n(M_0,z_{D})\,dM_0
\end{equation}
is the total number of descendant haloes in the relevant mass range.  For
sufficiently small $(M-m)$, $\Rz$ is simply related to the merger rate per
halo, $B/n$, by
\begin{equation}
	\Rzeqn\sim\int_{x}^{X}\!\! \frac{B}{n}\,d\xi\,.
\end{equation}

\subsection{Connection to EPS} 
\label{EPSConnection}

The merger rates determined from the Millennium simulation can be compared
to the analytic predictions of the Extended Press Schechter (EPS) formalism
(\citealt{1991ApJ...379..440B, 1993MNRAS.262..627L}).  To relate our per
halo merger rate $B/n$ to EPS, we begin with equation~(2.18) of
\citet{1993MNRAS.262..627L} for
\begin{equation}
  \frac{d^2p}{d\ln\Delta M^{LC}\, dt}(M_1^{LC} \rightarrow M_2^{LC} | t)  \,,
\label{eq:EPSrate}
\end{equation}
the probability that a halo of mass $M_1^{LC}$ will merge with another halo
of mass $\Delta M^{LC}=M_2^{LC}-M_1^{LC}$ in time interval $dt$.  Their
notation (which we denote with superscripts 'LC') is related to ours by
$M^{LC}_2\rightarrow M_0$, with $M^{LC}_1$ and $\Delta M^{LC}$ mapped to
our progenitor masses $M_1$ and $M_2$.  As we will see below, the order is
ambiguous due to an inconsistency in the EPS model that stems from the
assumption of binary mergers.  To relate $d^2p/d\ln\Delta M^{LC}/dt$ to
$B/n$, we first multiply it by $n(M_1^{LC})$, then convert the variables to
$(M_0,\xi)$ (see below), and finally divide by $n(M_0)$.

Before presenting the actual equation relating the two rates, we note two
caveats.  First, in order to compute an analytical merger rate from EPS we
must assume that mergers are binary and perfectly mass conserving, i.e.,
$M_0=M_1+M_2$ in our notation.  Neither assumption is strictly true in
numerical simulations, e.g., Table~\ref{table:NumProgs} and
Fig.~\ref{fig:NumProgs} show the distributions of the progenitor
multiplicity $N_p$.  We defer to Section~\ref{Tests} for a detailed
discussion of the tests that we have performed to quantify the binary
nature and the degree of merger mass conservation in the Millennium
simulation.

Second, the EPS rate in equation~(\ref{eq:EPSrate}) is not symmetric in the
progenitor masses $M^{LC}_1$ and $\Delta M^{LC}$, in contrast to our merger
rate $B_{MM'}$ in equation~(\ref{B1}), which is constructed to be symmetric
in the progenitor masses $M$ and $M'$.  We will therefore get different EPS
rates depending on if $M_1^{LC}$ is chosen to be the bigger or smaller
progenitor.  We will examine both options below: (A) $\xi=\Delta
M^{LC}/M^{LC}_1\leq 1$ and (B) $\xi=M^{LC}_1/\Delta M^{LC}\leq 1$.

With these caveats in mind, we find that the per halo merger rate $B/n$
corresponds to the following expression in the EPS model:
\begin{equation}
  \frac{B(M_0,\xi,z)}{n(M_0,z)} \leftrightarrow \sqrt\frac{2}{\pi} 
   \frac{d\delta_c}{dz} \frac{1}{\sigma(M')}
   \left|\frac{d\ln\sigma}{d\ln M}\right|_{M'}
    \left[ 1-\frac{\sigma^2(M_0)}{\sigma^2(M')} \right]^{-3/2}
\label{eqn:EPS}
\end{equation}
where $M'$ can be the smaller progenitor, i.e., $M'=M_2=M_0\xi/(1+\xi)$
(option A), or the larger progenitor, i.e., $M'=M_1=M_0/(1+\xi)$ (option
B).  The variable $\sigma^2(M)$ is the variance of the linear density field
smoothed with a window function containing mass $M$, and
$\delta_c(z)\propto 1/D(z)$ is the standard density threshold, with $D(z)$
being the linear growth factor.  Note that the exponential dependence at
the high mass end of the halo mass function has cancelled out on the right
hand side of equation~(\ref{eqn:EPS}).  Also note that both sides of
equation~(\ref{eqn:EPS}) are for merger rates \emph{per redshift} and
not per time.

We present our results for the merger rates determined from the Millennium
simulation in the next section and compare them to the two EPS predictions
in Section~\ref{eps}.

%%%%%%%%%%%%%%%%%% Fig 5 %%%%%%%%%%%%%%%%%%%%%
\begin{figure}
\centering
\includegraphics{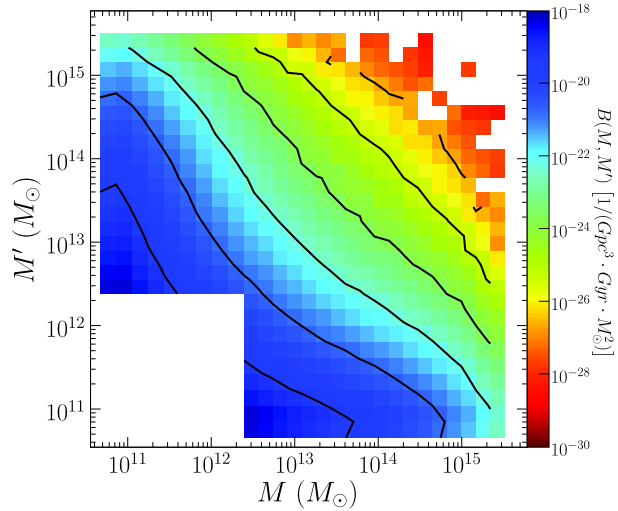}
\caption{Symmetric merger rate $B_{MM'}$ of equation~(\ref{B1}) as a
  function of progenitor masses $M$ and $M'$ computed from the $z$ = 0.06:0
  Millennium merger tree.  The merger rates decrease from blue to red; the
  overlaid black lines are contours of constant merger rates.}
\label{fig:Bmm} 
\end{figure}
%%%%%%%%%%%%%%%%%%%%%%%%%%%%%%%%%%%%%%%

\section{Results} 
\label{Results}

Throughout this section, we report our results from the Millennium
merger tree where the fragmented haloes are handled with the stitching
method.  We find the merger rates given by the snipping method to agree
with the stitching results to within 25\%.   Details of the comparison
are discussed in Section~\ref{SnipvsStitch}.

\subsection{Merger Rates at $z \approx 0$}
\label{MRz0}

%%%%%%%%%%%%%%%%%% Fig 6 %%%%%%%%%%%%%%%%%%%%%%
\begin{figure*}
\centering
\includegraphics{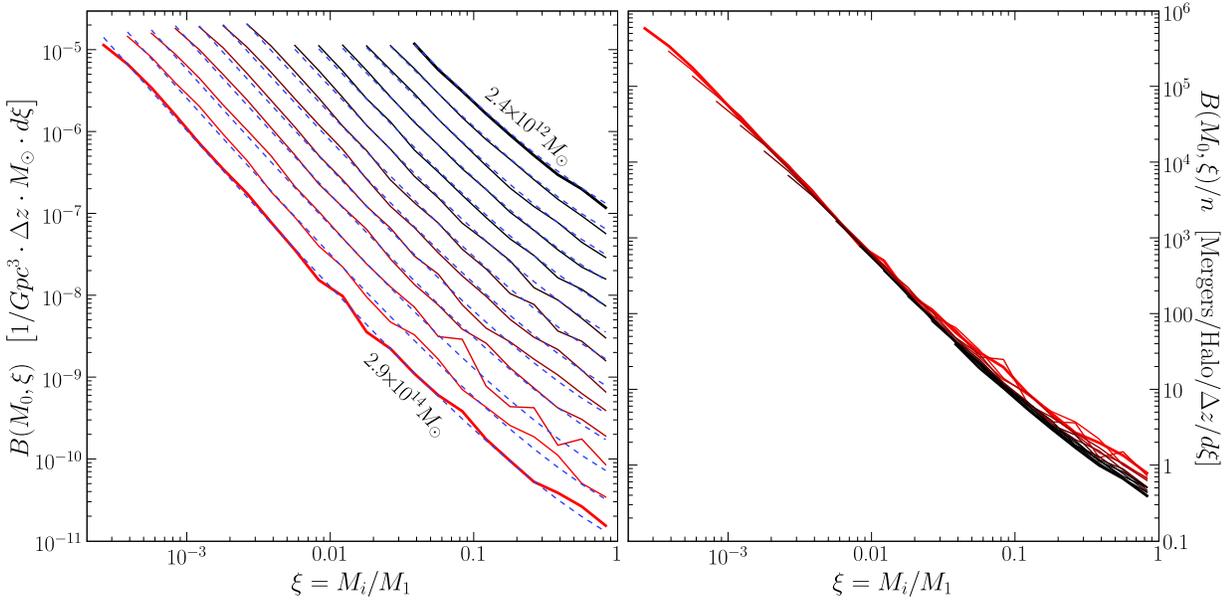}
\caption{Left panel: Mean merger rate $B(M_0,\xi)$ of equation~(\ref{B2})
  for the $z$ = 0.06:0 merger tree as a function of the mass ratio of the
  progenitors, $\xi$, for bins of fixed descendant halo mass $M_0$ (colour
  coded from black to red for increasing $M_0$).  The overlaid dashed blue
  lines are from our fitting formula in equation~(\ref{fiteqn}). Note that
  the presence of a fixed minimum mass resolution
  ($4.7\times10^{10}M_{\odot}$) corresponds to a minimum mass ratio $\xi$
  that decreases as $M_0$ increases.  Right panel: Mean merger rate {\it
    per halo}, $B(M_0,\xi)/n(M_0)$, of equation~(\ref{B3}) for the same
  tree.  Dividing out the halo number density $n(M_0)$ brings the curves on
  the left panel to nearly a single curve, indicating $B/n$ has very weak
  dependence on $M_0$.  }
\label{fig:B}
\end{figure*}
%%%%%%%%%%%%%%%%%%%%%%%%%%%%%%%%%%%%%%%%

Fig.~\ref{fig:Bmm} is a contour plot of the symmetric merger rate in
equation~(\ref{B1}), $B_{MM'}(M,M',\mt)$, calculated using the stitching
merger tree constructed from the $z$ = 0.06:0 Millennium outputs. Darker
(bluer) regions denote higher merger rates, which are concentrated in the
lower $(M, M')$ corner because there are more low mass haloes.  Minor
mergers (off-diagonal) are more common than major mergers (along the
diagonal).  The lower left corner is blank due to our lower cutoff on the
descendant FOF mass ($\sim1000$ particles; $M_0\ga\mmin$).  The noisy
nature of the upper right corner is due to limited merger statistics at
$\sim 10^{15} M_\odot$.

As we discussed in Section~\ref{stats}, instead of progenitor masses $M$
and $M'$, it is often more illuminating to study merger rates as a function
of the descendant FOF halo mass $M_0$ and the mass ratio $\xi$ of the
progenitors.  This is shown in Fig.~\ref{fig:B} for the same dataset as in
Fig.~\ref{fig:Bmm}.  The left panel plots the merger rate
$B(M_0,\xi,0.06\!:\!0)$ of equation~(\ref{B2}) against the progenitor mass
ratio $\xi$ for fixed bins of descendant FOF mass $M_0$.  We observe that
the merger rate $B(M_0,\xi)$ is a power-law in the progenitor mass ratio
$\xi$ when $\xi \la 0.1$ and shows an upturn in the major merger regime.
The power-law index is close to $-2$ and is nearly independent of the
descendant mass $M_0$.  More precise values are given in the fitting form
in equation~(\ref{fiteqn}) and Table~\ref{table:FitParms} below.

\begin{figure}
\centering
\includegraphics{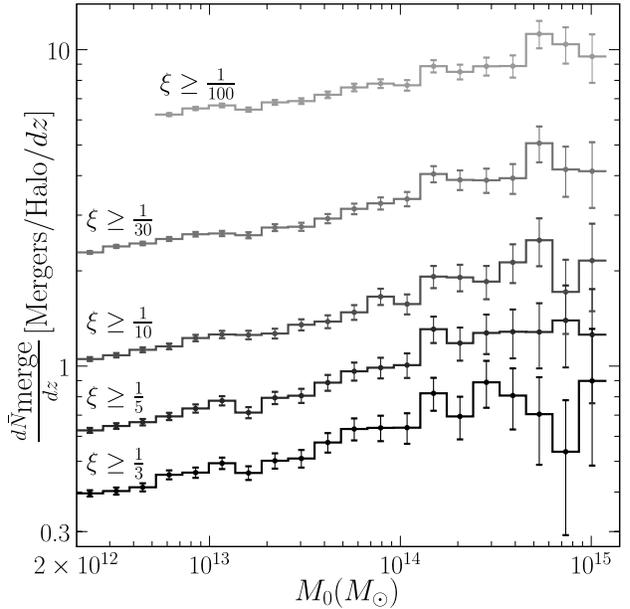}
\caption{Mean merger rate per halo (per unit $z$), $dN_{m}/dz$, as a
  function of descendant mass, $M_0$, for various ranges of the progenitor
  mass ratio $\xi$.  The upper curves include increasingly more minor
  mergers.  The $z$=0.06:0 merger tree is used.  Note the weak mass
  dependence over three decades of mass.  The error bars are computed
  assuming Poisson counting statistics in both the number of mergers and
  the number of haloes.}
\label{fig:Rmass}
\end{figure}

The main quantity we study in this paper is the mean merger rate {\it per
  descendant halo}, $B/n$, of equation~(\ref{B3}), shown in the right panel
of Fig.~\ref{fig:B}.  The rising amplitude of $B$ with decreasing $M_0$ is
remarkably largely removed when $B/n$ is plotted: the curves in the left
panel for different $M_0$ mass bins collapse onto nearly a single curve in
the right panel. This behaviour indicates that the merger rate per halo is
nearly independent of the descendant halo mass.  This weak mass dependence
is further illustrated in Fig.~\ref{fig:Rmass} and is also reported in
\cite{2007arXiv0708.1814G}.  As we will quantify in
Section~\ref{FitSection} below, the dependence on $M_0$ is approximately
$\propto M_0^{0.08}$.

Our lower cutoff of 40 particles for the progenitor FOF halo mass implies a
lower cutoff in the mass ratio of $\xi \geq
4.8\times10^{10}M_\odot/M_0$. This resolution cutoff is seen in the left
panel of Fig.~\ref{fig:B}, where we have sufficient halo statistics to
measure the merger rates for the higher mass haloes (lower curves) down to
very minor mergers, e.g., $\xi < 10^{-3}$ for $M_0 > 5 \times 10^{13}
M_\odot$; whereas the dynamic range is smaller for galaxy-size haloes,
e.g., $\xi > 0.01$ for $M_0 < 5\times 10^{12} M_\odot$.

The present-day merger rates shown in Fig.~\ref{fig:B} are all obtained
from the $z$ = 0.06:0 merger tree. The low-redshift outputs available from
the Millennium database in fact have a smaller spacing of $\Delta z \sim
0.02$.  We use the 0.06:0 merger tree to avoid any edge effects arising
from our stitching criterion that only subhalo fragments that remerge
within three outputs are stitched together (see Sec.~\ref{FOFLims}).  In
practice, this precaution is not critical and we find little difference
between the 0.06:0 and 0.02:0 results.

%%%%%%%%%%%%%%%%%%%%%%%%%%%%%%%%%%%%%%%
%%%%%%%%%%%%%%%%%%%%%%%%%%%%%%%%%%%%%%%

%%%%%%%%%%%%%%%%%%%%%%%%%%%%%%%%%%%%%%%
\begin{figure*}
\centering
\includegraphics{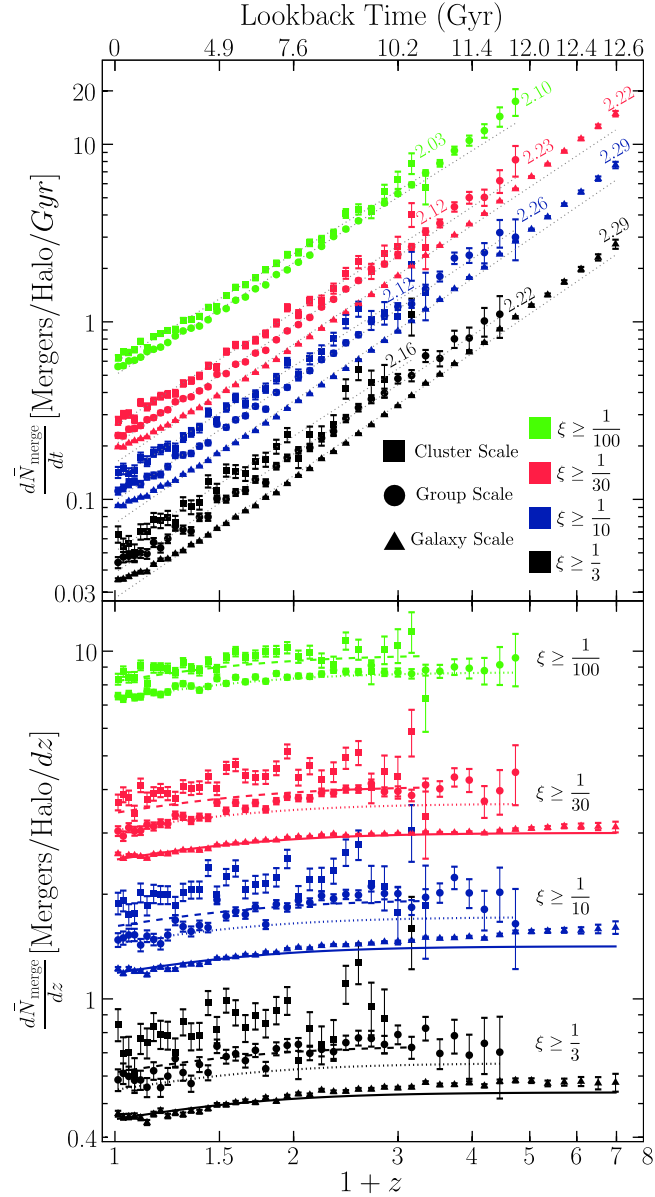}
\caption{Upper panel: Mean merger rate per halo (per Gyr), $dN_{m}/dt$, as a
  function of redshift for three bins of descendant mass $M_0$ and four
  ranges of progenitor mass ratio $\xi$ from the Millennium simulation
  (using the stitching tree).  The overlaid lines plot the best-fit
  power laws, $(1+z)^{n_m}$, with $n_m$ ranging from 2.03 to 2.29
  (labelled).  Note that power laws are reasonable fits at $z \ga 0.3$ but
  underpredict the Millennium rates at lower $z$.  Lower panel: Same as the
  upper panel but showing the merger rate $dN_{m}/dz$ per unit $z$ instead
  of per Gyr.  The dotted grey lines here show our fitting formula in
  equation~(\ref{fiteqn}), which is tuned to provide close fits at low $z$.
  In both panels, the error bars are computed assuming Poisson counting
  statistics in both the number of mergers and the number of haloes, and
  the curve for galaxy-scale haloes (triangles) with $\xi\geq\frac{1}{100}$
  (green) is suppressed because such minor mergers fall below the
  simulation resolution limit.  }
\label{fig:Rz}
\end{figure*}
%%%%%%%%%%%%%%%%%%%%%%%%%%%%%%%%%%%%%%%

\subsection{Merger Rates at Higher Redshift} 
\label{redshiftdependence}

Figs.~\ref{fig:B} and \ref{fig:Rmass} summarise our results for the $z=0$
merger rates.  At higher redshifts, the Millennium database provides
sufficient halo statistics for us to measure merger rates up to $z\sim 6$.
The results are shown in Fig.~\ref{fig:Rz}, where we plot the merger rate
per unit time (upper panel), $\R$, and per unit redshift (lower panel),
$\Rz$, as a function of redshift for three ranges of descendant masses
(galaxy, group, cluster) and four ranges of progenitor mass ratios ($\xi
\ge $ 1/3, 1/10, 1/30, and 1/100).  Errors are computed assuming Poisson
statistics for the number of mergers and haloes. We have suppressed merger
rates with poor merger statistics (and, therefore large error bars) to keep
the plots legible.

The mean merger rate per Gyr (upper panel) is seen to increase at higher
$z$.  We have fit power laws to each $M_0$ and $\xi$ range (dotted curves)
of the form
\begin{equation}
  \Reqn \propto(1+z)^{n_m}
\end{equation}
and find $n_m\sim 2-2.3$ for the ranges of $M_0$ and $\xi$ shown.  The
Millennium merger rates are seen to flatten out slightly at low $z$ and
deviate from a power law when the cosmological constant starts to dominate
the energy density of the universe.

A large number of merger rate statistics can be easily read off of
Fig.~\ref{fig:Rz}.  For example, at around $z=2$ ($z=4$) every FOF halo on
average experiences $\sim 2$-4 (10) minor mergers ($\xi \la 1/30$) per Gyr,
and about 10-20\% (70-90\%) of FOF haloes experience a major merger ($\xi
\ga 1/3$) every Gyr.

Unlike the rising $\R$, the merger rate {\it per unit redshift}, $\Rz$,
shows a remarkably weak dependence on $z$ in Fig.~\ref{fig:Rz} (lower
panel), increasing only slightly between $z=0$ and 1 and staying nearly
constant for $z\ga 1$ for all ranges of $M_0$ and $\xi$ shown.  The
overlaid curves are computed by integrating over the fitting form for $B/n$
to be discussed below (Sec.~\ref{FitSection}).

At $z>0$, Fig.~\ref{fig:Rz} shows that the dependence of $\Rz$ on
progenitor ratio $\xi$ and descendant mass $M_0$ is similar to the $z=0$
merger rates shown in Fig.~\ref{fig:B}: minor mergers occur more frequently
than major mergers, and the dependence on $M_0$ is weak, with galaxy-scale
haloes (triangles) on average experiencing fewer mergers (per halo) than
cluster-size haloes (squares).

\begin{table*}
\centering
\begin{tabular}{rcccccccc}
	Method & $A$ & $\tilde{\xi}$ & $\alpha$ & $\beta$ & $\gamma$ & $\eta$ & $\chi^2_\nu$ \\
	\hline
	Snip & 0.0101 & 0.017 & 0.089 & -2.17 & 0.316 & 0.325 & 1.86 \\
	Stitch & 0.0289 & 0.098 & 0.083 & -2.01 & 0.409 & 0.371 & 1.05 \\ 
\end{tabular}
\caption{Best fit parameters for equation~(\ref{fiteqn}).}
\label{table:FitParms}
\end{table*}

\subsection{A Universal Fitting Form} 
\label{FitSection}

We now propose a fitting form that can be used to approximate the halo
merger rates in the Millennium simulation discussed in the last two
subsections to an accuracy of 10-20\%.  The key feature we will use to
simplify the fit is the nearly universal form of the merger rate (per halo)
$B(M_0,\xi)/n$ shown in the right panel of Fig.~\ref{fig:B}, and the weak
redshift dependence shown in the bottom panel of Fig.~\ref{fig:Rz}.  We
find that the following functional form works well:
\begin{equation}
  \frac{B(M_0,\xi,z)}{n(M_0,z)} = A\left (\frac{M_0}{\tilde{M}}\right)^{\alpha}\xi^{\beta}
   \exp\left[\left(\frac{\xi}{\tilde{\xi}}\right)^{\gamma}\right]
    \left(\frac{d \delta_c}{dz}\right)^\eta \,,  
\label{fiteqn}
\end{equation}
where $\tilde{M}=1.2\times10^{12}M_{\odot}$ is a constant and
$\delta_c(z)\propto 1/D(z)$ is the standard density threshold normalised to
$\delta_c=1.686$ at $z=0$, with $D(z)$ being the linear growth factor.
Note that equation~(\ref{fiteqn}) is separable with respect to the three
major variables $M_0$, $\xi$, and $z$.

The form of the redshift dependence in equation~(\ref{fiteqn}) is chosen so
that $\eta=1$ corresponds to the EPS prediction in
equation~(\ref{eqn:EPS}).  In addition, this form has weak $z$ dependence
at $z\ga 1$ since the growth factor approaches that of the Einstein-de
Sitter model, $\delta_c(z)=1.68(1+z)$, and $d\delta_c/dz$ approaches a
constant.  This behaviour matches the weak redshift dependence seen in the
Millennium merger rate (bottom panel of Fig.~\ref{fig:Rz}).

To determine the parameters in equation~(\ref{fiteqn}), we fit
simultaneously to all redshifts $z<1$, mass ratios $\xi>10^{-3}$, and
masses $10^{12} \la M_0 \la 10^{14} M_\odot$.  The $B/n$ data points are
weighted using their Poisson distributed errors.  The resulting fits are
plotted as dotted curves in Figs.~\ref{fig:B} and \ref{fig:Rz}, and the fitting
parameters are given in Table~\ref{table:FitParms}, along with the overall
reduced $\chi^2_\nu$ obtained by fitting to all redshifts $z<1$
simultaneously. In addition to computing a global $\chi^2_\nu$ we also
compute a local $\chi^2_\nu(z)$ at each redshift and find relatively good
convergence across the $z<1$ redshift range: $\chi^2_\nu(z)$ remains below
1.5 for stitching and below 2 to 3 for snipping.

We note that the fitting form of equation~(\ref{fiteqn}) does not appear
symmetric in the progenitor masses $M_1$ and $M_2$ because by construction,
$\xi\equiv M_2/M_1<1$.  However, for any pair of progenitors, we identify
$M_1$ with the more massive and $M_2$ with the less massive progenitor and
then compute $\xi=M_2/M_1<1$.  This procedure yields the same $\xi$ and
therefore the same $B/n$ regardless of the order of the input progenitors,
in contrast to the EPS model discussed in Section~\ref{EPSConnection},
which is intrinsically asymmetric in $M_1$ and $M_2$.

\section{Tests} \label{Tests}

%%%%%%%%%%%%%%%%%%%%%%%%%%%%%%%%%%%%%%%
\begin{figure}
\centering
\includegraphics{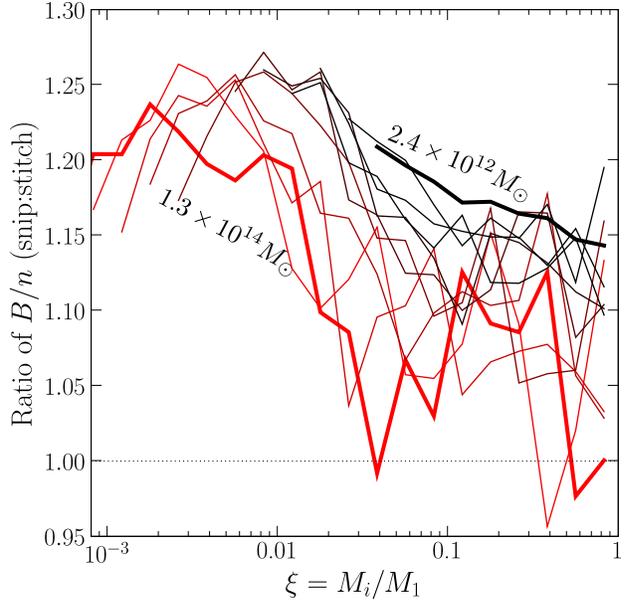}
\caption{The ratio of the snipping and stitching $B/n$ as a function of
  mass ratio $\xi$ computed using the 0.06:0 catalogue for a variety of mass
  bins in the range $2.4\times10^{12}M_\odot$ (black) $\leq M\leq$
  $1.3\times10^{14}$ (red). We find differences at the 25\% level at low
  $\xi$ with the snipping method consistently predicting a higher merger
  rate at all $\xi$.  We attribute this to the population of remerging
  orphan haloes.}
\label{fig:SnipStitchRatio}
\end{figure}
%%%%%%%%%%%%%%%%%%%%%%%%%%%%%%%%%%%%%%%

\subsection{Snipping vs Stitching Trees} \label{SnipvsStitch} 

Fig.~\ref{fig:SnipStitchRatio} shows the ratio of the $z=0$ per-halo merger
rate $B/n$ from the snipping and stitching methods.  Overall, the merger
rates given by the two methods differ by no more than 25\% over 2-3 orders
of magnitude in both the progenitor mass ratio $\xi$ and the descendant
mass $M_0$.  Within this difference, however,
Fig.~\ref{fig:SnipStitchRatio} and Table~\ref{table:FitParms} show that the
snipping method systematically yields a higher merger rate and a steeper
slope in the $\xi$-dependence than the stitching method.  These additional
merger events come from the orphaned subhaloes that are first snipped and
subsequently remerge (see Fig.~\ref{fig:fragmentation}).  Moreover, as
discussed in Section~\ref{FOFLims}, the fragmentation-to-merger ratio is
higher for more minor subhalo fragments (those with low fragment-to-FOF
mass ratios). There are therefore more remerging orphan haloes with lower
$\xi$, leading to the larger difference between snipping and stitching at
low $\xi$ seen in Fig.~\ref{fig:SnipStitchRatio}.

A remaining issue is our choice of the stitching criterion: as described in
Section~\ref{FOFLims}, we stitch only FOF fragments that are observed to
remerge within the next 3 outputs.  This choice is motivated by the fact
that about half of the halo fragments at a given output will have remerged
within three outputs (see Fig.~\ref{fig:fragdist}), and that such a small
$\dz$ criterion will allow us to effectively compute instantaneous merger
rates.  We have tested this criterion further by implementing a more
aggressive stitching algorithm that stitches \emph{all} fragments,
regardless of whether they eventually remerge.  We call this
$\infty$-stitching.  This algorithm represents the opposite limit to the
snipping method and may err on the side of {\it under-estimating} the
merger rates since it would stitch together close-encounter fly-by events
that do not result in actual mergers within a Hubble time.  We find the
amplitude of $B/n$ from $\infty$-stitching to be lower than that from the
3-stitching by up to $\sim 25$\%, similar in magnitude but opposite in sign
to the difference between snipping vs 3-stitching shown in
Fig.~\ref{fig:SnipStitchRatio}.  The fitting form in
equation~(\ref{fiteqn}) works well for $\infty$-stitching, where the best
fit parameters are
$A=0.0344,\tilde{\xi}=0.125,\alpha=0.118,\beta=-1.921,\gamma=0.399$, and
$\eta=0.853$.  This algorithm shows excellent convergence properties (see
\S\ref{TimeResolutionSection}) and excellent mass conservation properties
(\S\ref{MassCons}) but alters the FOF mass function by a few percent.

Since the snipping algorithm tends to inflate the merger rate and the
$\infty$-stitching algorithm tends to under-estimate it, we believe the
3-stitching used in all the results in Section~\ref{Results} should be a
fairly robust scheme.

\subsection{Convergence With Respect to $\dz$} 
\label{TimeResolutionSection}

%%%%%%%%%%%%%%%%%%%%%%%%%%%%%%%%%%%%%%%
\begin{figure*}
\centering
\includegraphics{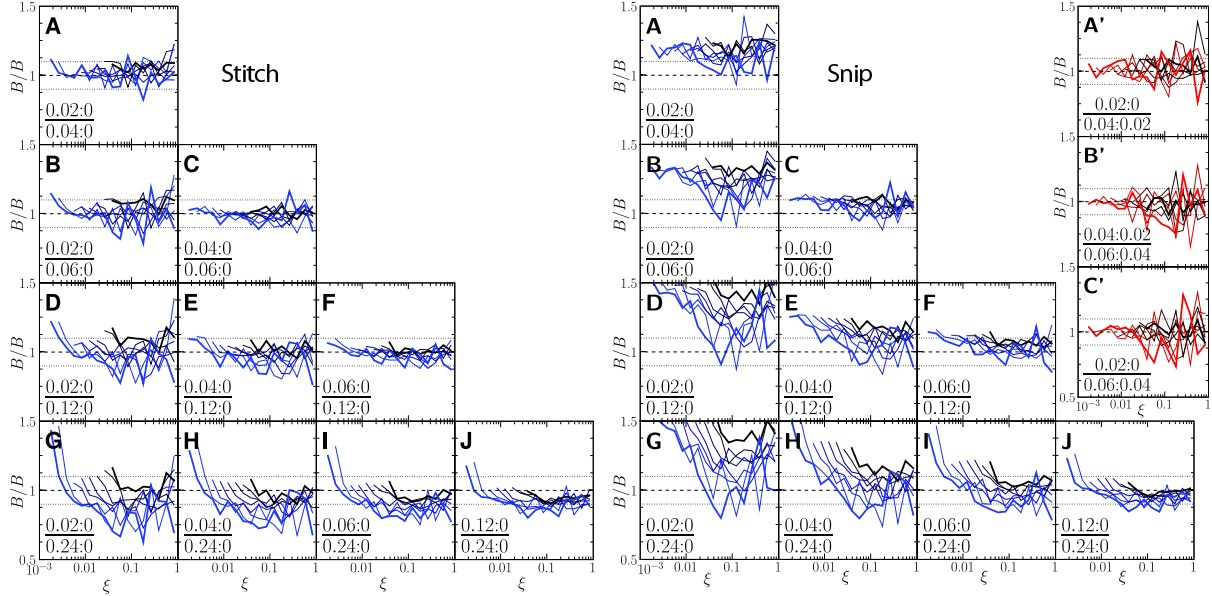}
\caption{$\dz$ Convergence Matrix (stitching left, snipping right). Each
  subplot is the ratio of $B/n$ for two different catalogues (labelled). The
  dashed lines denote equality and the dotted lines are the 10\% deviation
  levels. The ratios are presented for a variety of mass bins with the high
  mass bin highlighted in thick blue (or red) and the low mass bin
  highlighted in thick black.}
\label{fig:TR}
\end{figure*}
%%%%%%%%%%%%%%%%%%%%%%%%%%%%%%%%%%%%%%%

We have performed a number of tests to quantify the dependence of our
merger rate results on the choice of $\dz$ between the Millennium outputs
used to construct the merger trees.  It is not a priori clear which value
of $\dz$ is optimal: small $\dz$ can result in poor merger statistics since
most haloes would not have had time to merge; whereas large $\dz$ does not
have the time resolution to track individual merger events accurately and
also runs the risk of smearing out real redshift dependent effects.  The
optimal $\dz$ may also vary with redshift.

Our first test focuses on $z\approx 0$ mergers and quantifies how $B/n$
varies with the $\dz$ used to construct the trees.  Fig.~\ref{fig:TR} shows
the ratios of $B/n$ for five pairs of progenitor and descendant redshifts:
$(z_P,z_D)=$(0.02:0), (0.04:0), (0.06:0), (0.12:0), and (0.24:0),
corresponding to a time interval of $\Delta t=0.26, 0.54, 0.83, 1.44$, and
2.77 Gyr, respectively.  For the stitching method (left panel), there is
excellent convergence for $\dz\la 0.12$ (panels A-F), where the ratios of
$B/n$ are centred around 1 and rarely deviate beyond the 10\% level (dotted
line).  For $\dz=0.24$ (panels G-J), the ratios start to drop below unity.
This is consistent with the slowly rising merger rates with increasing $z$
shown in Fig.~\ref{fig:Rz}.  Thus, the stitching method yields merger trees
with robust $\dz$ convergence properties near $z=0$, and 
we have chosen $\dz=0.06$ to compute the merger rates in earlier sections.

The snipping method (right panel) shows inferior $\dz$ convergence.  The
$B/n$ computed with smaller $\dz$ consistently show higher merger rates
than those computed with larger $\dz$.  Moving up the left column (panels
G,D,B,A), we observe only some degree of convergence.  Better convergence
is seen along the main diagonal (panels A,C,F,J) in order of increasing
$\dz$.  In particular, panels C and F show excellent convergence properties
(to the 10\% level) centred around (0.06:0).  To emphasise that the problem
is with $\dz$ and not with a particular output (say, any possible edge
effects at $z$ = 0 or 0.02), we show in panels A',B',C' the ratios of $B/n$
computed using three merger trees with the same $\dz=0.02$ but centred at
progressively higher $z$: $z$ = (0.02:0), (0.04:0.02), and (0.06:0.04).
The agreement is excellent, in striking contrast to panel B.  Based on
these tests, we have chosen to use $\dz=0.06$ for the snipping method.

We believe that the snipping method has inferior $\dz$ convergence
properties because of the remerging orphan subhaloes (see
Section~\ref{FOFLims} and Fig.~\ref{fig:fragmentation}).  These
fragmentation events are sewn together in the stitching scheme and
therefore do not contribute to the merger rates. In the snipping scheme,
however, the snipped events provide a fresh supply of haloes, many of which
remerge in the next few outputs.  This effect artificially boosts the
merger rate across small $\dz$.

\begin{figure}
\centering
\includegraphics{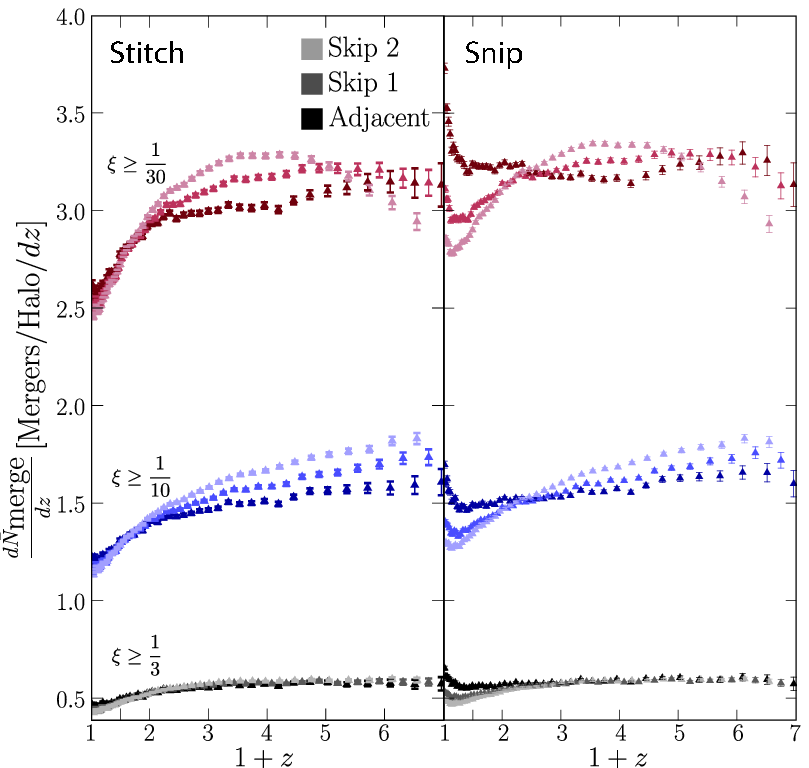}
\caption{ Merger rate $\Rz$ computed using three types of redshift
  spacings: \emph{adjacent}, \emph{skip 1} and \emph{skip 2} (see text).
  For clarity, only the galaxy-mass haloes are shown; the group and cluster
  haloes behave similarly.}
\label{fig:RzTR}
\end{figure}

Our second $\dz$ convergence test is performed at all redshifts.  We test
three types of spacing: (1) {\it Adjacent} spacing uses adjacent
catalogues, e.g., at low $z$, it uses (0.02:0), (0.04:0.02), (0.06:0.04);
(2) \emph{Skip 1} spacing skips an output, e.g., (0.04:0), (0.06:0.02)....,
and (3) \emph{Skip 2} spacing skips two outputs, e.g., (0.06:0),
(0.09:0.02), and so on.  Fig.~\ref{fig:RzTR} shows $\Rz$ computed using
these three $\dz$ for galaxy-mass haloes.  We again see excellent
$\dz$-convergence for the stitching method at $z\la 1.5$ (left panel) and
worse $\dz$-convergence for the snipping method (right panel).  The latter
follows the behaviour seen in Fig~\ref{fig:TR}, with adjacent spacing
($\dz=0.02$ at $z=0$) over-predicting the merger rate.

At higher redshifts ($z\ga1.5$), Fig.~\ref{fig:RzTR} shows that the merger
rates in the minor merger regime differ by up to $\sim 15$\% depending on
which of the three types of spacing is used.  This difference is not likely
to be due to the fragmentation events since as Fig.~\ref{fig:fragdist}
shows, the ratio of fragmentation to merger events is 40\% near $z=0$ but
drops to $\la$10\% for $z\ga3$.  Rather, we believe that the inferior $\dz$
convergence at high $z$ is due to the increasing $\dz$ between Millennium
outputs (e.g., the smallest $\dz$ is $\sim 0.5$ at $z\sim6$ vs $\dz=0.02$
at $z\approx 0$) and the inaccuracy of the multiple counting ordering
assumption for large $\dz$ (see Section~\ref{MultiCountValid}).  At high
$z$, we therefore advocate using the finest output spacings available in
the Millennium database, noting the good time convergence for major mergers
($\xi\ga1/3$) but $\sim 15$\% variations in the minor merger rates.

%%%%%%%%%%%%%%%%%%% Fig 11 %%%%%%%%%%%%%%%%%%%%
%%%%%%%%%%%%%%%%%%%%%%%%%%%%%%%%%%%%%%%

\subsection{Multiple vs Binary Counting} \label{MultiCountValid}

%%%%%%%%%%%%%%%%%%% Fig 12 %%%%%%%%%%%%%%%%%%%%
\begin{figure*}
\centering
\includegraphics[width=6.5in]{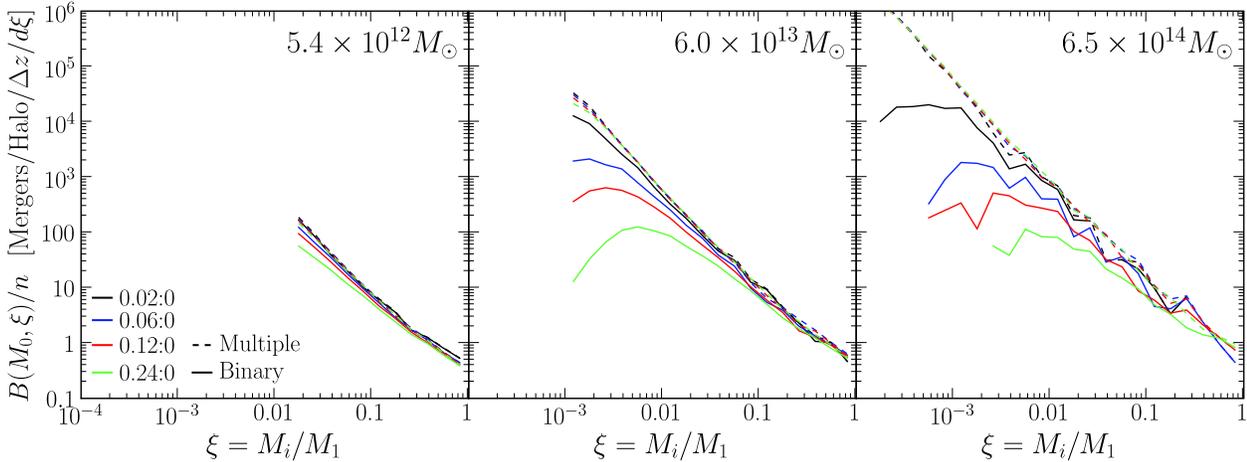}
\caption{Comparisons of merger rate per halo, $B/n$, computed via multiple
  counting (dashed lines) and binary counting (solid lines) for 4 merger
  trees with increasing $\Delta z$.  Three descendant mass bins are shown
  (from left to right).  We find the multiple counting rate to be in
  excellent agreement regardless of $\Delta z$ of the tree, whereas the
  binary counting $B/n$ curves fall off from the observed power-law behaviour
  towards lower $\xi$.}
\label{fig:Binary}
\end{figure*}
%%%%%%%%%%%%%%%%%%%%%%%%%%%%%%%%%%%%%%%

As discussed in Section~\ref{MultiCount}, for descendant FOF haloes with
more than two progenitor haloes, we include all progenitors when we
calculate the merger rates for completeness.  Since mergers are often
assumed to be binary events, we have tested the difference between our
multiple counting results and those obtained by counting only the two most
massive progenitors of a given descendant halo. Fig.~\ref{fig:Binary}
compares the merger rates (per halo), $B(M_0,\xi)/n$, for these two
counting methods (dashed: multiple; solid: binary) as a function of the
progenitor mass ratio $\xi$ for three descendant masses $M_0$ (increasing
from left to right).  In each panel, results from four merger trees using
increasing $\dz$ of 0.02, 0.06, 0.12, and 0.24 are shown.

The most notable trend in Fig.~\ref{fig:Binary} is that the multiple
counting method gives similar merger rates regardless of $\dz$, indicating
good time convergence in the results (as we have discussed in detail in
Section~\ref{TimeResolutionSection}).  The binary counting rates, on the
other hand, deviate increasingly from the multiple rates when larger $\dz$
are used because the binary assumption becomes less valid for larger
$\dz$. As a function of $\xi$, the binary and multiple merger rates match
well in the major merger regime but deviate significantly for small $\xi$.
This occurs because binary counting counts only the two most massive
progenitors and ignores the additional (typically low-mass) progenitors.
It therefore closely approximates the major-merger rates but
under-estimates the minor-merger regime of the multiple counting result.

Fig.~\ref{fig:Binary} suggests that for a given minimum mass resolution
(i.e. a minimum $\xi$), there is a corresponding $\Delta z$ for which the
binary counting method is a good approximation.  For example, for $6\times
10^{13} M_\odot$ haloes (centre panel), the binary and multiple merger
rates are similar down to $\xi \approx 0.05$ for $\Delta z\sim 0.12$, and
down to $\xi \approx 0.005$ when $\Delta z$ is decreased to 0.02.  Thus the
multiple counting $B/n$ can be thought of as the small-$\dz$ limit of the
binary $B/n$.

Another test we have performed is on the ordering of mergers assumed in the
multiple counting method described in Section~\ref{MultiCount}.  There, we
assumed that the less massive progenitors $M_2,M_3,...$ each merged with
the most massive progenitor $M_1$ and not with one another.  This
assumption is motivated by the fact that satellite haloes in $N$-body
cosmological simulations are typically seen to accrete onto a much more
massive host halo as a minor merger event instead of merging with another
satellite halo.  We have quantified the validity of this assumption by
taking large $\dz$ in the Millennium outputs, applying this ordering, and
checking against the actual merging order among the progenitors when finer
$\dz$ is used.  (Of course, we cannot do this for the minimum $\dz=0.02$
available in the database.)  The fraction of misordering naturally rises
with increasing $\dz$ due to the degraded time resolution, but for $\dz \la
0.06$, we find the fraction of progenitors to have merged with a progenitor
other than $M_1$ to be $\la 10\%$.  Most of the mergers among multiple
progenitors, therefore, do occur between the most massive progenitor and a
less massive progenitor, as we have assumed.

%%%%%%%%%% Fig 13 %%%%%%%%%%%%%%%%%%%%%%%%%%%%%
\begin{figure}
\centering
\includegraphics{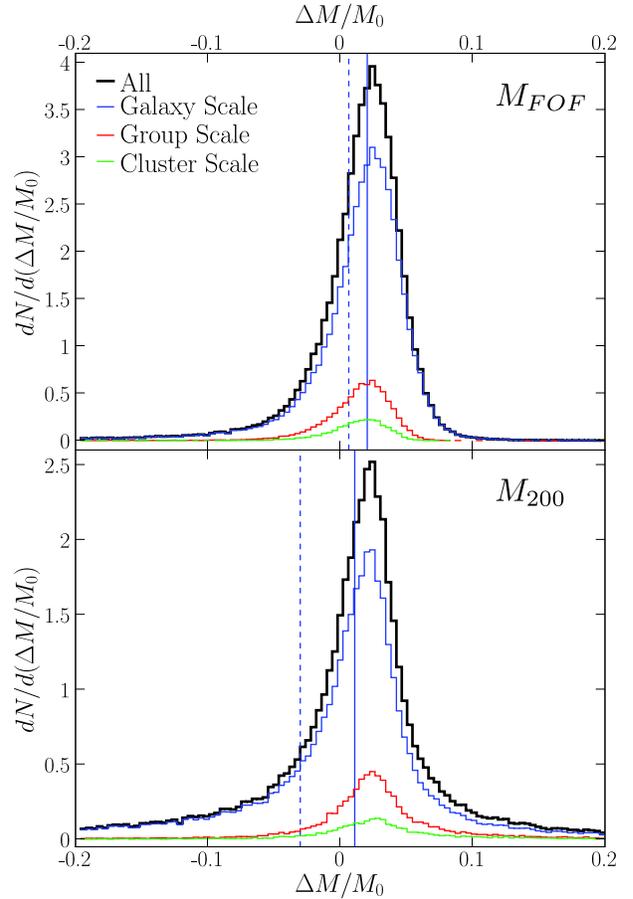}
\caption{Distributions of $\Delta M/M_0$ from the $z$ = 0.06:0 Millennium
  merger tree (using stitching; snipping is nearly identical) computed
  using the $\mfof$ mass (top panel) vs $\mvir$ virial mass (bottom
  panel).  The solid vertical line is the median of the distribution for
  the galaxy-mass bin and the dashed line is the mean.  We note a longer
  negative $\Delta M$ tail for the $\mvir$ tree when compared to the
  $\mfof$ tree.  Note, however, that the peaks of the two distributions are
  in good agreement ($\Delta M/M_0 \sim 2.5\%$)}
\label{fig:DM}
\end{figure}
%%%%%%%%%%%%%%%%%%%%%%%%%%%%%%%%%%%%%%%%%%

\subsection{Mass Conservation and  ``Diffuse'' Accretion} 
\label{MassCons}

Thus far we have analysed mergers in terms of the progenitor halo mass
$M_i$ and the descendant halo mass $M_0$.  Mergers are, however, messy
events, and the sum of $M_i$ does not necessarily equal $M_0$.  To quantify
this effect, we define a ``diffuse'' component, $\Delta M$, for a given
descendant halo:
\begin{equation}
	M_0=\sum_{i=1}^{N_p} M_i + \Delta M\,,
\end{equation}
where $\Delta M$ is diffuse in the sense that it is not resolved as
distinct haloes in the simulation.  A non-zero $\Delta M$ can be due to
physical processes such as tidal stripping and diffuse mass accretion that
cause a net loss or gain in halo mass after a merger event.  In
simulations, additional numerical factors also contribute to $\Delta M$ due
to different algorithms used in, for example, defining halo mass (FOF vs
spherical overdensity).  $\Delta M$ therefore does not necessarily have to
be positive in every merger event.

Fig.~\ref{fig:DM} shows the distribution of $\Delta M/M_0$ for the
$z=$(0.06:0) Millennium merger tree.  Only haloes that have experienced
mergers between these two redshifts (i.e. those with more than one
progenitor) are plotted.  The snipping tree (not shown) gives a very
similar distribution as the stitching tree shown here.  For comparison, we
have computed $\Delta M/M_0$ using the two different halo mass definitions
$M_{FOF}$ and $\mvir$.  The distribution shows a prominent peak at $\Delta
M/M_0 \sim 2.5$ \% for both $M_{FOF}$ and $M_{200}$, indicating that in the
majority of the merger events between $z=0.06$ and 0.0, $\sim 97.5$\% of
the mass of the descendant halo comes from resolvable progenitor haloes.

Even though the two distributions in Fig.~\ref{fig:DM} have similar peaks,
the $\mvir$ mass definition produces longer $\Delta M/M_0$ tails than the
$\mfof$ mass definition, and the mean of the distribution (dotted vertical
line) is negative for $\mvir$.  We believe this is because mass definitions
based on the assumption of spherical symmetry (as $\mvir$ does) have
difficulties assigning accurate mass to non-spherical FOF haloes and tend
to underestimate the halo mass (see, e.g., \citealt{2001A&A...367...27W}).
$M_{FOF}$, on the other hand, can account for all the mass in a given FOF
object that is identified as 'merged' by the FOF halo finder well before
virialization.  As discussed in Section~\ref{DarkMatterHaloes}, we have
been using the $\mfof$ mass thus far.

Our main results on merger rates in Section~\ref{Results} were determined
for numerically resolved dark matter haloes; they are therefore not
affected directly by the fact that $\Delta M$ is generally non-zero for
merger events.  We find, however, that $\Delta M/M_0$ increases with
$\Delta z$, and this diffuse accretion component makes an important
contribution to the mass growth of a halo over its lifetime.  We will
explore the growth of haloes in further detail in subsequent papers.

\subsection{Halo Mass Function} 
\label{MassFunction}

%%%%%%%%%%%%%%%% Fig 14 %%%%%%%%%%%%%%%% 
\begin{figure}
\centering
\includegraphics{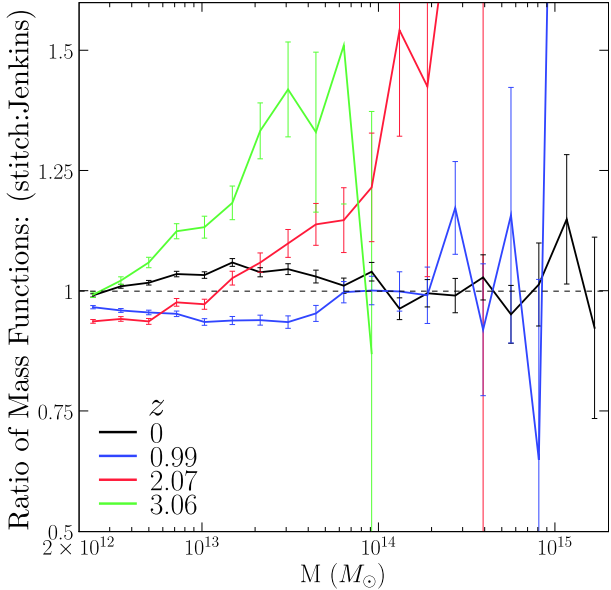}
\caption{Ratios of the Millennium halo mass function (computed from the
  stitching trees) to the fit of \citet{2001MNRAS.321..372J} using the
  $\mfof$ mass.  The results for the snipping trees are virtually
  identical.  We note a significant deviation of $\sim 25$\% at $z\sim 3$
  between the Jenkins fit and the Millennium mass function.}
\label{fig:MassFunction}
\end{figure}
%%%%%%%%%%%%%%%%%%%%%%%%%%%%%%%%%%%%%%%%%

The mass function of dark matter haloes in principle depends on both the
definition of halo mass and the algorithm used to treat fragmentation
events.  As illustrated in Fig.~\ref{fig:fragmentation}, the snipping
method by construction preserves the original FOF mass function, while the
stitching scheme modifies it slightly as it rearranges fragmented subhaloes
between FOFs.  We find that the impact on the mass function is negligible
(less than $\sim 0.25$\%) so will use the stitching result below.

Fig.~\ref{fig:MassFunction} shows the ratio of the Millennium halo mass
function to the fits by \citet{2001MNRAS.321..372J} for $\mfof$ at four
redshifts: $z\approx 0$, 1, 2, and 3.  The fit of Jenkins et al is accurate
to better than 10\% for low redshift ($z\la 1$), but it underestimates the
Millennium halo abundance by $\ga 25$\% at the high mass end for $z>1$.
This discrepancy is present but not obvious on the log-log plot in Fig.~2
of \cite{2005Natur.435..629S}.  \cite{2007astro.ph..2360L} also noted this
difference.  Since the stitching and snipping mass functions are virtually
identical, this appears to be a discrepancy between the Millennium FOF
catalogue and the fit of \citet{2001MNRAS.321..372J}.

\section{Theoretical Models for Merger Rates} 
\label{DiscussionSection}

\subsection{Extended Press-Schechter Model} 
\label{eps}

\begin{figure}
\centering
\includegraphics{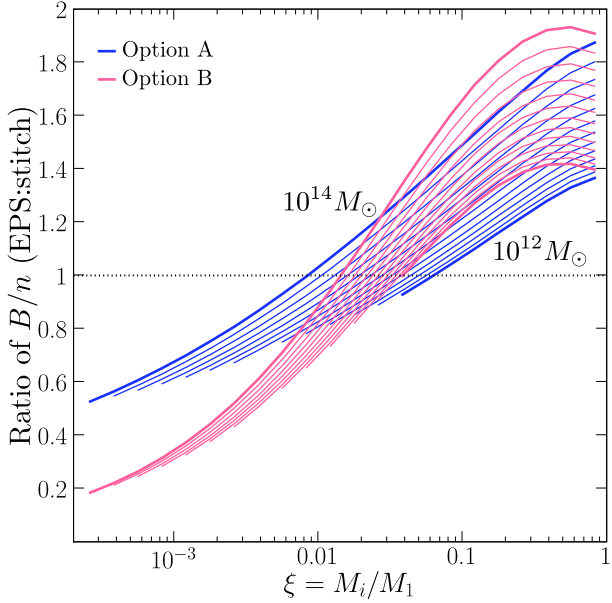}
\caption{Comparison between our Millennium merger rate (from the fit) and
  the two predictions of the Extended Press-Schechter model.  The ratio of
  $B/n$ from EPS to Millennium is plotted.  Blue and red label the two
  options in assigning progenitor masses in the EPS model (see text);
  within each colour, the set of curves from bottom to top denotes
  increasing $M_0$ bins, from $\sim10^{12}M_\odot$ to
  $\sim3\times10^{14}M_\odot$.  The EPS model is seen to overpredict the
  major merger rate by up to a factor of $\sim 2$ and underpredicts the
  minor merger rate by up to a factor of $\sim 5$.}
\label{fig:EPS}
\end{figure}

In Section~\ref{EPSConnection} we discussed how our merger rates are
related to the conditional probabilities in the EPS model and obtained
equation~(\ref{eqn:EPS}), where there are two choices for the definition of
the progenitor mass $M'$ since the EPS model is not symmetric in the two
progenitor masses.  In Fig.~\ref{fig:EPS} we show the ratio of the EPS
prediction to our Millennium $B/n$ at $z=0$, where we have computed the EPS
rates given by the right-hand-side of equation~(\ref{eqn:EPS}) using the
same cosmological parameters as for the Millennium simulation.  We compute
the variance of the smoothed linear density field, $\sigma^2(M)$, in the
$\Lambda$CDM cosmology using the power spectrum fit provided in
\cite{1999ApJ...511....5E}.

Fig.~\ref{fig:EPS} shows that EPS {\it underpredicts} the $z=0$ rate for
minor mergers by up to a factor of $\sim 5$, and {\it overpredicts} the
rate at $\xi \ga 0.05$, indicating that the dependence of the EPS merger
rate on $\xi$ is shallower than our $B/n \sim \xi^\beta$, where the
best-fit $\beta$ is $-2.17$ and $-2.01$ for the snipping and stitching
methods, respectively (see Table~\ref{table:FitParms}).  In terms of the
descendant mass $M_0$, the dependence of the EPS rate is too steep compared
to our $B/n$, leading to the spread in each bundle of curves in
Fig.~\ref{fig:EPS}.  The two choices of $M'$ in EPS are seen to lead to
different predictions.  Assigning $M'$ to be the smaller progenitor (option
A) results in a somewhat smaller discrepancy than option B.

Fig.~\ref{fig:EPS} compares the rates at $z=0$.  At higher redshifts, we
find the Millennium merger rate to evolve as $\propto (d\delta_c/dz)^\eta$,
where $\eta \approx 0.37$ (see equation~[\ref{fiteqn}] and
Table~\ref{table:FitParms}) and is shallower than the EPS prediction of
$\eta=1$ in equation~(\ref{eqn:EPS}).  Since the functional forms of both
our fit for $B/n$ and the EPS expression are separable with respect to
$M_0$, $\xi$ and $z$, the $z=0$ curves in Fig.~\ref{fig:EPS} will maintain
the same shape at higher $z$, but the amplitude of the ratio will increase.
For instance, the ratio shown in Fig.~\ref{fig:EPS} will be increased by a
factor of 1.26, 1.32, 1.34, and 1.35 at $z=1$, 2, 4, and 6 respectively.
The discrepancy between the Millennium results and the EPS predictions is
therefore even worse at higher $z$.

Given that the Press-Schechter mass function is known not to match the halo
abundances in simulations very closely, it is not particularly surprising
that the EPS merger rates in Fig.~\ref{fig:EPS} do not match the Millennium
results closely.  The substantial discrepancy, however, does highlight the
limitation of the EPS model and provides the motivation to build more
accurate merger rates based on improved PS mass functions.  We address this
issue in separate papers \citep{2008arXiv0801.3459Z}, in which we
investigate a moving density-barrier algorithm to generate merger trees
that produces a better match to simulation results than the constant
barrier of the PS model.

%%%%%%%%%%%%%%%% Fig 15 %%%%%%%%%%%%%%%% 
%%%%%%%%%%%%%%%%%%%%%%%%%%%%%%%% 

\subsection{Halo Coagulation} \label{Coagulation}

The merging of dark matter haloes is, in principle, a coagulation
process. Coagulation is often modelled by the Smoluchowski coagulation
equation \citep{Smoluchowski}, which governs the time evolution of the mass
function $n(M,t)$ of the objects of interest with a coagulation kernel.  In
the absence of fragmentations, the time change of $n$ is given by
\begin{equation}\label{eqn:coag}
\begin{split}
  \frac{dn(M)}{dt}=\frac{1}{2}\int_0^M\!\!\!\! A(M',M\!\!-\!\! M') n(M')
  n(M\!\!-\!\! M') dM' \\- \int_0^\infty \!\!\!\! A(M,M') n(M') n(M) dM'  \,,
\end{split}
\end{equation}
where the first term on the right-hand side is a source term due to mergers
of two smaller haloes of mass $M'$ and $M-M'$, while the second term is a
sink term due to haloes in the mass bin of interest merging with another
halo of mass $M'$, forming a halo of higher mass $M+M'$.  When applied to
hierarchical structure formation, $A(M,M')$, the symmetric coagulation
kernel (in units of volume/time), tracks the probability for a halo of mass
$M$ to merge with a halo of mass $M'$.  Our merger rate per halo, $B/n$,
can be simply related to $A$ by
\begin{equation}\label{eqn:Btocoag}
	A(M,M') \leftrightarrow \frac{B(M,M')}{n(M)n(M')} \,.
\end{equation}
We note, however, that the coagulation equation in the form of
equation~(\ref{eqn:coag}) is valid only for mass-conserving binary
mergers. As seen throughout this paper, these assumptions are not strictly
true in numerical simulations, and modifications are required to account
for the issues that have been discussed thus far, such as net mass gain or
loss (i.e. $\Delta M \neq 0$), multiple merger events, and halo
fragmentation. While the relative errors may be small when integrated over
a small time interval, repeated application of equation~(\ref{eqn:coag})
using equation~(\ref{eqn:Btocoag}) may not yield robust results.

\begin{figure}
\centering
\includegraphics{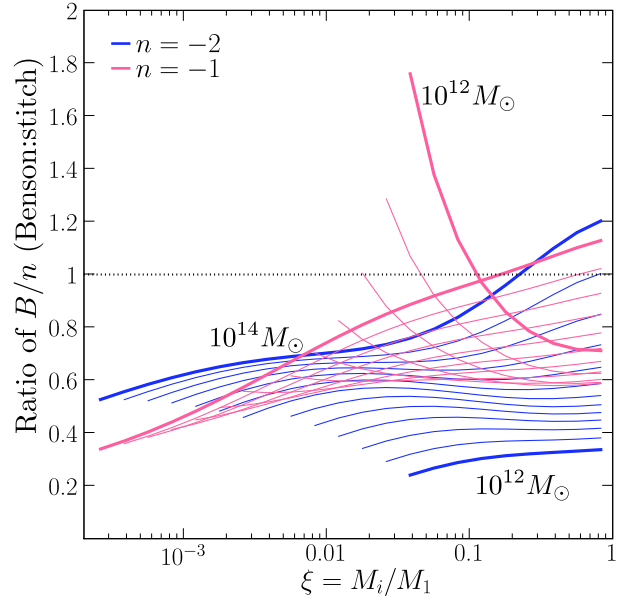}
\caption{Same as Fig.~\ref{fig:EPS}, only now comparing the merger rates
  from Benson et al. (2005) to the Millennium $B/n$ for initial power
  spectrum index $n=-2$ (blue) and $n=-1$ (red).}
\label{fig:Benson}
\end{figure}

Assuming that $n(M)$ is the Press-Schechter mass function,
\cite{2005MNRAS.357..847B} have developed numerical techniques to construct
the coagulation kernel for self-similar cosmological models with initial
power-law power spectrum $P(k)\propto k^n$.  Their technique is
underconstrained and does not yield a unique expression for $A(M,M')$.  In
order to pick out a particular solution, a regularisation condition was
applied to force $A(M,M')$ to vary smoothly.  We have transformed the
coordinates of their fits to $A(M,M')$ to compare their results with our
merger rate $B/n$.  Fig.~\ref{fig:Benson} shows the ratio of their fitting
formula to the Millennium $\Lambda$CDM merger rate for spectral indices
$n=-1$ and $-2$ as a function of progenitor mass ratio $\xi$ for various
descendant halo mass bins.  The difference can be up to a factor of
several.

%%%%%%%%%%%%%%%% Fig 16 %%%%%%%%%%%%%%%% 
%%%%%%%%%%%%%%%%%%%%%%%%%%%%%%%%%%%%% 

\section{Conclusions and Discussions}

In this paper we have computed the merger rates of dark matter FOF haloes as a
function of descendant halo mass $M_0$, progenitor mass ratio $\xi$, and
redshift $z$ using the merger trees that we constructed from the halo
catalogue of the Millennium simulation.  Our main results are presented in
Figs.~\ref{fig:B} to \ref{fig:Rz}, which show very simple and nearly
separable dependence on $M_0$, $\xi$, and $z$. The mean merger rate per
descendant FOF halo, $B/n$, is seen to depend very weakly on the halo mass
$M_0$ (Fig.~\ref{fig:B} right panel and Fig.~\ref{fig:Rmass}). As a
function of redshift $z$, the per halo merger rate in units of per Gyr
increases as $(1+z)^\alpha$, where $\alpha\sim 2$ to 2.3 (top panel of
Fig.~\ref{fig:Rz}), but when expressed in units of per redshift, the merger
rate depends very weakly on $z$ (bottom panel of
Fig.~\ref{fig:Rz}). Regardless of $M_0$ and $z$, the dependence of $B/n$ on
the progenitor mass ratio, $\xi = M_i/M_1$, is a power law to a good
approximation in the minor merger regime ($\xi \la 0.1$) and shows an
upturn in the major merger regime (Fig.~\ref{fig:B}). These simple
behaviours have allowed us to propose a universal fitting formula in
equation~(\ref{fiteqn}) that is valid for $10^{12}\leq M_0\la 10^{15}
M_\odot$, $\xi \ga 10^{-3}$, and up to $z\sim 6$.

Throughout the paper we have emphasised and quantified the effects on the
merger rates due to events in which a progenitor halo fragments into
multiple descendant haloes.  We have shown that the method commonly used to
remove these fragmented haloes in merger trees -- the snipping method --
has relatively poor $\dz$-convergence (Figs.~\ref{fig:TR} and
\ref{fig:RzTR}).  Our alternative approach -- the stitching method --
performs well with regards to this issue without drastically modifying the
mass conservation properties or the mass function of the Millennium FOF
catalogue (Figs.~ \ref{fig:DM} and \ref{fig:MassFunction}).

We have computed the two predictions for merger rates from the analytical
EPS model for the same $\Lambda$CDM model used in the Millennium
simulation.  At $z=0$, we find the EPS major merger rates to be too high by
50-100\% (depending on halo mass) and the minor merger rates to be too low
by up to a factor of 2-5 (Fig.~\ref{fig:EPS}).  The discrepancy increases
at higher $z$.

The coagulation equation offers an alternative theoretical framework for
modelling the mergers of dark matter haloes.  We have discussed how our
merger rate is related to the coagulation merger kernel in theory.  In
practice, however, we find that mergers in simulations are not always
mass-conserving binary events, as assumed in the standard coagulation form
given by equation~(\ref{eqn:coag}).  Equation~(\ref{eqn:coag}) will
therefore have to be modified before it can be used to model mergers in
simulations.

\cite{2001ApJ...546..223G} studied the rate of major mergers (defined to be
$\xi \ge 1/3$ in our notation) in $N$-body simulations and found a steeper
power law dependence of $\propto (1+z)^3$ (at $z\la 2$) for the merger rate
per Gyr than ours.  Their simulations did not have sufficient mass
resolution to determine the rate at $z \ga 2$.  It is important to note,
however, that our $B/n$ at redshift $z$ measures the instantaneous rate of
mergers during a small $\Delta z$ interval at that redshift.  By contrast,
they studied the merging history of {\it present-day} haloes and measured
only the rate of major mergers for the most massive progenitor at redshift
$z$ of a $z=0$ halo (see their paragraph 4, section 2).  A detailed
comparison is outside the interest of this paper.

Mergers of dark matter haloes are related to but not identical to mergers
of galaxies.  It typically takes the stellar component of an infalling
galaxy extra time to merge with a central galaxy in a group or cluster
after their respective dark matter haloes have been tagged as merged by the
FOF algorithm.  This time delay is governed by the dynamical friction
timescale for the galaxies to lose orbital energy and momentum, and it
depends on the mass ratios of the galaxies and the orbital parameters
(\citealt{2008MNRAS.383...93B} and references therein).  In addition to
this difference in merger timescale, the growth in the stellar mass of a
galaxy is not always proportional to the growth in its dark matter halo
mass.  A recent analysis of the galaxy catalogue in the Millennium
simulation \citep{2007arXiv0708.1814G} finds galaxy growth via major
mergers to depend strongly on stellar mass, where mergers are more
important in the buildup of stellar masses in massive galaxies while star
formation is more important in galaxies smaller than the Milky Way.
Extending the analysis of this paper to the mergers of {\it subhaloes} in
the Millennium simulation will provide the essential link between their and
our results.  

For similar reasons, our results for the evolution of the dark matter halo
merger rate per Gyr ($(1+z)^{n_m}$ with $n_m\sim2-2.3$) cannot be trivially
connected to the observed merger rate of \emph{galaxies}.  It is
nonetheless interesting to note that a broad disagreement persists in the
observational literature of galaxy merger rates.  The reported power law
indices $n_m$ have ranged from 0 to 5 (see, for example,
\citealt{1994ApJ...429L..13B, 1994ApJ...435..540C, 1995ApJ...445...37Y,
  1995ApJ...454...32W, 1997ApJ...475...29P, 2000MNRAS.311..565L,
  2002ApJ...565..208P,2003AJ....126.1183C,
  2004ApJ...601L.123B,2004ApJ...612..679L, 2004ApJ...617L...9L}).
\cite{2006ApJ...652...56B} followed the redshift evolution of
subhalo mergers in N-body simulations and provided a more detailed comparison
with recent observations by, e.g., \cite{2004ApJ...617L...9L} that find
$n_m<1$.  They attributed such a weak redshift evolution in the number
of close companions per galaxy to the fact that the high merger rate per
halo at early times is counteracted by a decrease in the number of haloes
massive enough to host a galaxy pair.

The merger rates in this paper are global averages over all halo
environments.  The rich statistics in the Millennium simulation allow for
an in-depth analysis of the environmental dependence of dark matter halo
merger rates, which we will report in a subsequent paper (Fakhouri \& Ma,
in preparation).

\section*{Acknowledgements}

We have enjoyed enlightening discussions with Mike Boylan-Kolchin, Liang
Gao, Ari Laor, Simon White, Andrew Zentner, and Jun Zhang.  This work is
supported in part by NSF grant AST 0407351.  The Millennium Simulation
databases used in this paper and the web application providing online
access to them were constructed as part of the activities of the German
Astrophysical Virtual Observatory.

\section*{Appendix: The Durham Tree}

In this paper we have used two methods to handle fragmentation events in
the Millennium FOF merger trees: snipping and stitching.  Here we discuss
and compare a third method used by the Durham group
\citep{2006MNRAS.370..645B, 2006MNRAS.367.1039H, 2003MNRAS.338..903H}.

The Durham algorithm is designed to reduce spurious linkings of FOF haloes
in low-density regions.  Before constructing the FOF merger tree, they
filter through the Millennium FOF and subhalo database, and split up a
subhalo from its FOF halo if (1) the subhalo's centre is outside twice the
half mass radius of the FOF halo, or (2) the subhalo has retained more than
75\% of the mass it had at the last output time at which it was an
independent halo \citep{2006MNRAS.367.1039H}.  Condition (1) is effectively
a spatial cut, while (2) is based on the argument that less massive
subhaloes are expected to undergo significant stripping as they merge with
more massive haloes.  This algorithm then \emph{discards} the subhaloes that
are split off from FOF groups at $z=0$, along with any associated
progenitor subhaloes.  Around 15\% of the original FOF haloes are split in
this algorithm.

The Durham algorithm tends to reduce the number of fragmented haloes in
the resulting trees, but it does not eliminate all such events.  A method
much like our snipping method is used to treat the remaining fragmentation
events.  The resulting FOF tree is available at the Millennium public
database along with the original Millennium tree.

\begin{figure}
\centering
\includegraphics{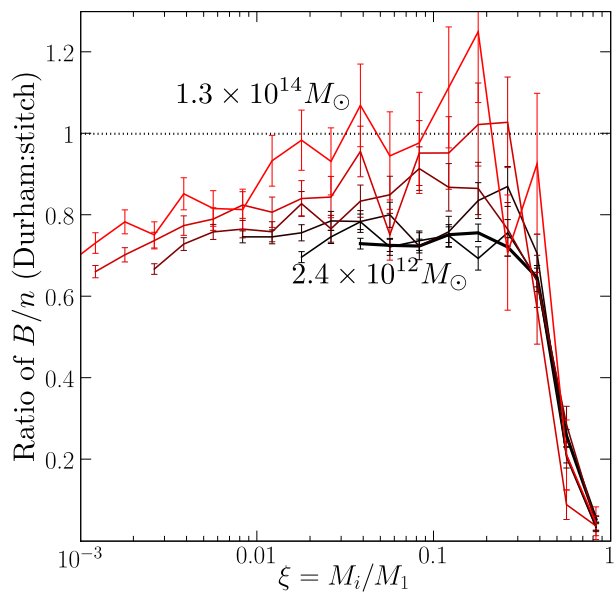}
\caption{The ratio of the Durham merger rate $B/n$ to our stitching rate
  $B/n$ (section~\ref{FitSection}) as a function of progenitor mass ratio
  $\xi$ for a number of descendant mass bins ranging from
  $\sim2\times10^{12}M_\odot$ (black) to $\sim 10^{14} M_\odot$ (red). The
  Durham merger rate tends to be lower than the stitching merger rate, and
  suffers a sudden drop in the major merger regime ($\xi\ga0.3$).}
\label{fig:DurhamB}
\end{figure}

To compare with our stitching and snipping trees, we have repeated all of
our merger rate calculations and tests using the Durham
tree. Fig.~\ref{fig:DurhamB} shows the ratio of the resulting Durham merger
rate, $B/n$, to that from our stitching tree at $z=0$.  The Durham rate is
generally lower than our rate for minor mergers (by up to $\sim 30$\%), and
it drops precipitously for major mergers ($\xi \ga 0.3$).  The two
additional conditions applied to split up subhaloes in the Durham algorithm
therefore appear to have eliminated most of the major merger events.

\begin{figure}
\centering
\includegraphics{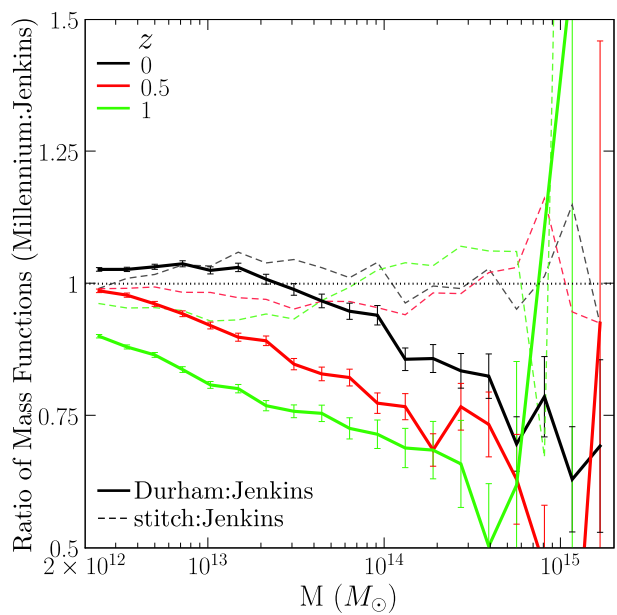}
\caption{The ratio of the Durham halo mass function to the fit of Jenkins
  et al. (2001) at redshifts 0, 0.5, and 1 (thick solid curves with error
  bars).  The ratio of the halo mass function from our stitching method to
  the same fit is shown for comparison (thin dashed curves).  The Durham
  algorithm tends to reduce the masses of massive haloes, leading to a
  deficit that grows to $\sim 50\%$ at $\sim 10^{15} M_\odot$ and at higher
  $z$.}
\label{fig:DurhamMFunc}
\end{figure}

Moreover, these splitting conditions in the Durham algorithm also modify
the halo mass function.  Fig.~\ref{fig:DurhamMFunc} shows the ratio of the
Durham mass function to the fit of Jenkins et al. (2001) at $z=0$, 0.5, and
1 (thick solid curves with error bars).  The ratio of our stitching mass
function to the same fit is overlaid for comparison (thin dotted
curves).  The Durham mass function is systematically lower: the
number of $z=0$ haloes with $M\ga 10^{14} M_\odot$ is $\sim 25$\% lower,
and the difference increases at $z\sim 1$, affecting the halo mass function
even at $M\sim 2\times 10^{12} M_\odot$.

We believe that the deficit of major merger events and massive haloes in
the Durham catalogue is partially due to their second criterion that
splits off subhaloes that have retained 75\% of their original mass.  This
condition may indeed remove spurious FOF linkings in the minor merger
regime, but major merger events tend to preserve much of the original
progenitor masses and have been systematically split by the Durham
algorithm.

\begin{figure}
\centering
\includegraphics{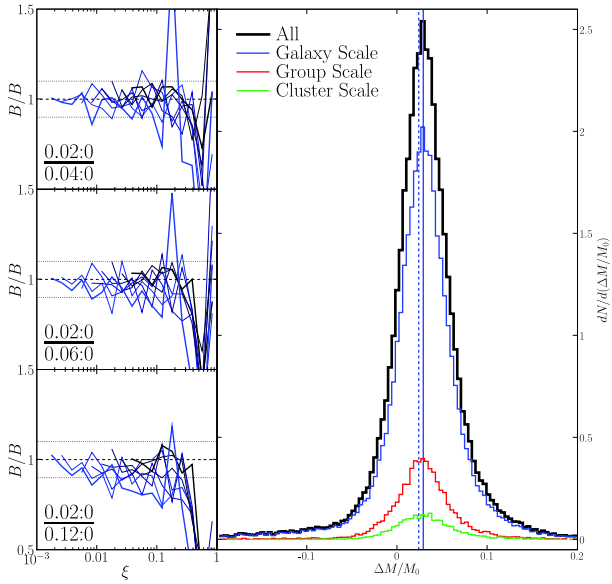}
\caption{Left panels: A subset of the $\dz$ convergence matrix presented in
  Fig.~\ref{fig:TR}, now computed using the Durham tree.  Note the poor
  convergence properties of the major merger end ($\xi\ga 0.3$).  This
  corresponds to the region of the largest difference between the
  stitching and Durham merger rates (see Fig.~\ref{fig:DurhamB}).  Right
  panel: The distribution of $\Delta M/M_0$ for the $z$ = 0.06:0 Durham catalogue
  (similar to Fig.~\ref{fig:DM}).}
\label{fig:DurhamTRDM}
\end{figure}

Finally, Fig.~\ref{fig:DurhamTRDM} (right panel) shows that the Durham tree
has similar mass conservation properties as our stitching tree in
Fig.~\ref{fig:DM}.  The distribution of the mass in the 'diffuse'
component, $\Delta M/M_0$, has a very similar peak of $\sim 3$\%, although
the negative $\Delta M$ events have been suppressed. For $\dz$ convergence
(left panels; cf section~\ref{TimeResolutionSection}), the Durham tree
performs well in the minor merger regime but is consistently poor for major
mergers, again probably due to the splitting condition (2) above.

The Durham algorithm is tuned to address questions of galaxy evolution.
The issues we have uncovered regarding this algorithm are specifically for
the mergers of dark matter haloes, the subject of this paper; issues with
the mergers of galaxies will require a separate study.

\bibliographystyle{mn2e}
\bibliography{Mergers}

\begin{thebibliography}{}

\bibitem[\protect\citeauthoryear{{Benson}, {Cole}, {Frenk}, {Baugh} \&
  {Lacey}}{{Benson} et~al.}{2000}]{2000MNRAS.311..793B}
{Benson} A.~J.,  {Cole} S.,  {Frenk} C.~S.,  {Baugh} C.~M.,    {Lacey} C.~G.,
  2000, MNRAS, 311, 793

\bibitem[\protect\citeauthoryear{{Benson}, {Kamionkowski} \&
  {Hassani}}{{Benson} et~al.}{2005}]{2005MNRAS.357..847B}
{Benson} A.~J.,  {Kamionkowski} M.,    {Hassani} S.~H.,  2005, MNRAS, 357, 847

\bibitem[\protect\citeauthoryear{{Berrier}, {Bullock}, {Barton}, {Guenther},
  {Zentner} \& {Wechsler}}{{Berrier} et~al.}{2006}]{2006ApJ...652...56B}
{Berrier} J.~C.,  {Bullock} J.~S.,  {Barton} E.~J.,  {Guenther} H.~D.,
  {Zentner} A.~R.,    {Wechsler} R.~H.,  2006, ApJ, 652, 56

\bibitem[\protect\citeauthoryear{{Bond}, {Cole}, {Efstathiou} \&
  {Kaiser}}{{Bond} et~al.}{1991}]{1991ApJ...379..440B}
{Bond} J.~R.,  {Cole} S.,  {Efstathiou} G.,    {Kaiser} N.,  1991, ApJ, 379,
  440

\bibitem[\protect\citeauthoryear{{Bower}, {Benson}, {Malbon}, {Helly}, {Frenk},
  {Baugh}, {Cole} \& {Lacey}}{{Bower} et~al.}{2006}]{2006MNRAS.370..645B}
{Bower} R.~G.,  {Benson} A.~J.,  {Malbon} R.,  {Helly} J.~C.,  {Frenk} C.~S.,
  {Baugh} C.~M.,  {Cole} S.,    {Lacey} C.~G.,  2006, MNRAS, 370, 645

\bibitem[\protect\citeauthoryear{{Boylan-Kolchin}, {Ma} \&
  {Quataert}}{{Boylan-Kolchin} et~al.}{2008}]{2008MNRAS.383...93B}
{Boylan-Kolchin} M.,  {Ma} C.-P.,    {Quataert} E.,  2008, MNRAS, 383, 93

\bibitem[\protect\citeauthoryear{{Bundy}, {Fukugita}, {Ellis}, {Kodama} \&
  {Conselice}}{{Bundy} et~al.}{2004}]{2004ApJ...601L.123B}
{Bundy} K.,  {Fukugita} M.,  {Ellis} R.~S.,  {Kodama} T.,    {Conselice} C.~J.,
   2004, ApJL, 601, L123

\bibitem[\protect\citeauthoryear{{Burkey}, {Keel}, {Windhorst} \&
  {Franklin}}{{Burkey} et~al.}{1994}]{1994ApJ...429L..13B}
{Burkey} J.~M.,  {Keel} W.~C.,  {Windhorst} R.~A.,    {Franklin} B.~E.,  1994,
  ApJ, 429, L13

\bibitem[\protect\citeauthoryear{{Carlberg}, {Pritchet} \&
  {Infante}}{{Carlberg} et~al.}{1994}]{1994ApJ...435..540C}
{Carlberg} R.~G.,  {Pritchet} C.~J.,    {Infante} L.,  1994, ApJ, 435, 540

\bibitem[\protect\citeauthoryear{{Cole}, {Lacey}, {Baugh} \& {Frenk}}{{Cole}
  et~al.}{2000}]{2000MNRAS.319..168C}
{Cole} S.,  {Lacey} C.~G.,  {Baugh} C.~M.,    {Frenk} C.~S.,  2000, MNRAS, 319,
  168

\bibitem[\protect\citeauthoryear{{Conselice}, {Bershady}, {Dickinson} \&
  {Papovich}}{{Conselice} et~al.}{2003}]{2003AJ....126.1183C}
{Conselice} C.~J.,  {Bershady} M.~A.,  {Dickinson} M.,    {Papovich} C.,  2003,
  AJ, 126, 1183

\bibitem[\protect\citeauthoryear{{Davis}, {Efstathiou}, {Frenk} \&
  {White}}{{Davis} et~al.}{1985}]{1985ApJ...292..371D}
{Davis} M.,  {Efstathiou} G.,  {Frenk} C.~S.,    {White} S.~D.~M.,  1985, ApJ,
  292, 371

\bibitem[\protect\citeauthoryear{{De Lucia}, {Springel}, {White}, {Croton} \&
  {Kauffmann}}{{De Lucia} et~al.}{2006}]{2006MNRAS.366..499D}
{De Lucia} G.,  {Springel} V.,  {White} S.~D.~M.,  {Croton} D.,    {Kauffmann}
  G.,  2006, MNRAS, 366, 499

\bibitem[\protect\citeauthoryear{{Eisenstein} \& {Hu}}{{Eisenstein} \&
  {Hu}}{1999}]{1999ApJ...511....5E}
{Eisenstein} D.~J.,  {Hu} W.,  1999, ApJ, 511, 5

\bibitem[\protect\citeauthoryear{{Gottl{\"o}ber}, {Klypin} \&
  {Kravtsov}}{{Gottl{\"o}ber} et~al.}{2001}]{2001ApJ...546..223G}
{Gottl{\"o}ber} S.,  {Klypin} A.,    {Kravtsov} A.~V.,  2001, ApJ, 546, 223

\bibitem[\protect\citeauthoryear{{Governato}, {Gardner}, {Stadel}, {Quinn} \&
  {Lake}}{{Governato} et~al.}{1999}]{1999AJ....117.1651G}
{Governato} F.,  {Gardner} J.~P.,  {Stadel} J.,  {Quinn} T.,    {Lake} G.,
  1999, AJ, 117, 1651

\bibitem[\protect\citeauthoryear{{Guo} \& {White}}{{Guo} \&
  {White}}{2007}]{2007arXiv0708.1814G}
{Guo} Q.,  {White} S.~D.~M.,  2007, ArXiv e-prints 0708.1814

\bibitem[\protect\citeauthoryear{{Harker}, {Cole}, {Helly}, {Frenk} \&
  {Jenkins}}{{Harker} et~al.}{2006}]{2006MNRAS.367.1039H}
{Harker} G.,  {Cole} S.,  {Helly} J.,  {Frenk} C.,    {Jenkins} A.,  2006,
  MNRAS, 367, 1039

\bibitem[\protect\citeauthoryear{{Helly}, {Cole}, {Frenk}, {Baugh}, {Benson} \&
  {Lacey}}{{Helly} et~al.}{2003}]{2003MNRAS.338..903H}
{Helly} J.~C.,  {Cole} S.,  {Frenk} C.~S.,  {Baugh} C.~M.,  {Benson} A.,
  {Lacey} C.,  2003, MNRAS, 338, 903

\bibitem[\protect\citeauthoryear{{Jenkins}, {Frenk}, {White}, {Colberg},
  {Cole}, {Evrard}, {Couchman} \& {Yoshida}}{{Jenkins}
  et~al.}{2001}]{2001MNRAS.321..372J}
{Jenkins} A.,  {Frenk} C.~S.,  {White} S.~D.~M.,  {Colberg} J.~M.,  {Cole} S.,
  {Evrard} A.~E.,  {Couchman} H.~M.~P.,    {Yoshida} N.,  2001, MNRAS, 321, 372

\bibitem[\protect\citeauthoryear{{Kang}, {Jing}, {Mo} \& {B{\"o}rner}}{{Kang}
  et~al.}{2005}]{2005ApJ...631...21K}
{Kang} X.,  {Jing} Y.~P.,  {Mo} H.~J.,    {B{\"o}rner} G.,  2005, ApJ, 631, 21

\bibitem[\protect\citeauthoryear{{Kauffmann}, {Colberg}, {Diaferio} \&
  {White}}{{Kauffmann} et~al.}{1999}]{1999MNRAS.303..188K}
{Kauffmann} G.,  {Colberg} J.~M.,  {Diaferio} A.,    {White} S.~D.~M.,  1999,
  MNRAS, 303, 188

\bibitem[\protect\citeauthoryear{{Kauffmann}, {White} \&
  {Guiderdoni}}{{Kauffmann} et~al.}{1993}]{1993MNRAS.264..201K}
{Kauffmann} G.,  {White} S.~D.~M.,    {Guiderdoni} B.,  1993, MNRAS, 264, 201

\bibitem[\protect\citeauthoryear{{Lacey} \& {Cole}}{{Lacey} \&
  {Cole}}{1993}]{1993MNRAS.262..627L}
{Lacey} C.,  {Cole} S.,  1993, MNRAS, 262, 627

\bibitem[\protect\citeauthoryear{{Lavery}, {Remijan}, {Charmandaris}, {Hayes}
  \& {Ring}}{{Lavery} et~al.}{2004}]{2004ApJ...612..679L}
{Lavery} R.~J.,  {Remijan} A.,  {Charmandaris} V.,  {Hayes} R.~D.,    {Ring}
  A.~A.,  2004, ApJ, 612, 679

\bibitem[\protect\citeauthoryear{{Le F{\`e}vre}, {Abraham}, {Lilly}, {Ellis},
  {Brinchmann}, {Schade}, {Tresse}, {Colless}, {Crampton}, {Glazebrook},
  {Hammer} \& {Broadhurst}}{{Le F{\`e}vre} et~al.}{2000}]{2000MNRAS.311..565L}
{Le F{\`e}vre} O.,  {Abraham} R.,  {Lilly} S.~J.,  {Ellis} R.~S.,  {Brinchmann}
  J.,  {Schade} D.,  {Tresse} L.,  {Colless} M.,  {Crampton} D.,  {Glazebrook}
  K.,  {Hammer} F.,    {Broadhurst} T.,  2000, MNRAS, 311, 565

\bibitem[\protect\citeauthoryear{{Lin}, {Koo}, {Willmer}, {Patton},
  {Conselice}, {Yan}, {Coil}, {Cooper}, {Davis}, {Faber}, {Gerke},
  {Guhathakurta} \& {Newman}}{{Lin} et~al.}{2004}]{2004ApJ...617L...9L}
{Lin} L.,  {Koo} D.~C.,  {Willmer} C.~N.~A.,  {Patton} D.~R.,  {Conselice}
  C.~J.,  {Yan} R.,  {Coil} A.~L.,  {Cooper} M.~C.,  {Davis} M.,  {Faber}
  S.~M.,  {Gerke} B.~F.,  {Guhathakurta} P.,    {Newman} J.~A.,  2004, ApJl,
  617, L9

\bibitem[\protect\citeauthoryear{{Lukic}, {Heitmann}, {Habib}, {Bashinsky} \&
  {Ricker}}{{Lukic} et~al.}{2007}]{2007astro.ph..2360L}
{Lukic} Z.,  {Heitmann} K.,  {Habib} S.,  {Bashinsky} S.,    {Ricker} P.~M.,
  2007, ArXiv e-prints 0702360

\bibitem[\protect\citeauthoryear{{Maller}, {Katz}, {Kere{\v s}}, {Dav{\'e}} \&
  {Weinberg}}{{Maller} et~al.}{2006}]{2006ApJ...647..763M}
{Maller} A.~H.,  {Katz} N.,  {Kere{\v s}} D.,  {Dav{\'e}} R.,    {Weinberg}
  D.~H.,  2006, ApJ, 647, 763

\bibitem[\protect\citeauthoryear{{Murali}, {Katz}, {Hernquist}, {Weinberg} \&
  {Dav{\'e}}}{{Murali} et~al.}{2002}]{2002ApJ...571....1M}
{Murali} C.,  {Katz} N.,  {Hernquist} L.,  {Weinberg} D.~H.,    {Dav{\'e}} R.,
  2002, ApJ, 571, 1

\bibitem[\protect\citeauthoryear{{Patton}, {Pritchet}, {Carlberg}, {Marzke},
  {Yee}, {Hall}, {Lin}, {Morris}, {Sawicki}, {Shepherd} \& {Wirth}}{{Patton}
  et~al.}{2002}]{2002ApJ...565..208P}
{Patton} D.~R.,  {Pritchet} C.~J.,  {Carlberg} R.~G.,  {Marzke} R.~O.,  {Yee}
  H.~K.~C.,  {Hall} P.~B.,  {Lin} H.,  {Morris} S.~L.,  {Sawicki} M.,
  {Shepherd} C.~W.,    {Wirth} G.~D.,  2002, ApJ, 565, 208

\bibitem[\protect\citeauthoryear{{Patton}, {Pritchet}, {Yee}, {Ellingson} \&
  {Carlberg}}{{Patton} et~al.}{1997}]{1997ApJ...475...29P}
{Patton} D.~R.,  {Pritchet} C.~J.,  {Yee} H.~K.~C.,  {Ellingson} E.,
  {Carlberg} R.~G.,  1997, ApJ, 475, 29

\bibitem[\protect\citeauthoryear{{Sheth} \& {Tormen}}{{Sheth} \&
  {Tormen}}{2002}]{2002MNRAS.329...61S}
{Sheth} R.~K.,  {Tormen} G.,  2002, MNRAS, 329, 61

\bibitem[\protect\citeauthoryear{{Smoluchowski}}{{Smoluchowski}}{1916}]{Smoluc%
howski}
{Smoluchowski} M.,  1916, Phys. Zeit., 17, 557

\bibitem[\protect\citeauthoryear{{Somerville} \& {Primack}}{{Somerville} \&
  {Primack}}{1999}]{1999MNRAS.310.1087S}
{Somerville} R.~S.,  {Primack} J.~R.,  1999, MNRAS, 310, 1087

\bibitem[\protect\citeauthoryear{{Springel}, {White}, {Jenkins}, {Frenk},
  {Yoshida}, {Gao}, {Navarro}, {Thacker}, {Croton}, {Helly}, {Peacock}, {Cole},
  {Thomas}, {Couchman}, {Evrard}, {Colberg} \& {Pearce}}{{Springel}
  et~al.}{2005}]{2005Natur.435..629S}
{Springel} V.,  {White} S.~D.~M.,  {Jenkins} A.,  {Frenk} C.~S.,  {Yoshida} N.,
   {Gao} L.,  {Navarro} J.,  {Thacker} R.,  {Croton} D.,  {Helly} J.,
  {Peacock} J.~A.,  {Cole} S.,  {Thomas} P.,  {Couchman} H.,  {Evrard} A.,
  {Colberg} J.,    {Pearce} F.,  2005, Nat, 435, 629

\bibitem[\protect\citeauthoryear{{Springel}, {White}, {Tormen} \&
  {Kauffmann}}{{Springel} et~al.}{2001}]{2001MNRAS.328..726S}
{Springel} V.,  {White} S.~D.~M.,  {Tormen} G.,    {Kauffmann} G.,  2001,
  MNRAS, 328, 726

\bibitem[\protect\citeauthoryear{{White}}{{White}}{2001}]{2001A&A...367...27W}
{White} M.,  2001, A\&A, 367, 27

\bibitem[\protect\citeauthoryear{{Woods}, {Fahlman} \& {Richer}}{{Woods}
  et~al.}{1995}]{1995ApJ...454...32W}
{Woods} D.,  {Fahlman} G.~G.,    {Richer} H.~B.,  1995, ApJ, 454, 32

\bibitem[\protect\citeauthoryear{{Yee} \& {Ellingson}}{{Yee} \&
  {Ellingson}}{1995}]{1995ApJ...445...37Y}
{Yee} H.~K.~C.,  {Ellingson} E.,  1995, ApJ, 445, 37

\bibitem[\protect\citeauthoryear{{Zhang}, {Ma} \& {Fakhouri}}{{Zhang}
  et~al.}{2008}]{2008arXiv0801.3459Z}
{Zhang} J.,  {Ma} C.-P.,    {Fakhouri} O.,  2008, ArXiv e-prints, 801

\end{thebibliography}

\label{lastpage}

\end{document}